\documentclass[preprint]{elsarticle}
\usepackage{amsmath,graphicx}
\usepackage{doi}

\usepackage{color}
\usepackage{placeins}
\usepackage{float}
\usepackage{bm}
\usepackage{tabularx,colortbl}
\usepackage{url}

\usepackage{algorithm}
\usepackage[noend]{algpseudocode}

\usepackage{pgfplots}
\usepgfplotslibrary{fillbetween}
\usepackage{amssymb}
\pgfplotsset{compat=1.9}
\usepackage{comment}

\usepackage{tikz}
\usetikzlibrary{intersections}
\usetikzlibrary{shapes.geometric}
\usetikzlibrary{angles}
\usepackage{tikz-qtree}
\usetikzlibrary{fit}
\usetikzlibrary{spy,calc}
\usepackage{subfig}

\biboptions{sort&compress}

\newcommand{\R}{\mathbb{R}}

\pgfplotsset{select coords between index/.style 2 args={
    x filter/.code={
        \ifnum\coordindex<#1\fi
        \ifnum\coordindex>#2\fi
    }
}}

\pgfplotsset{filter discard warning=false}

\newenvironment{customlegend}[1][]{%
        \begingroup
        \csname pgfplots@init@cleared@structures\endcsname
        \pgfplotsset{#1}%
    }{%
        \csname pgfplots@createlegend\endcsname
        \endgroup
    }%
    \def\addlegendimage{\csname pgfplots@addlegendimage\endcsname}

\begin{document}
\begin{frontmatter}

\title{Image Compression with Isotropic and Anisotropic 
    Shepard Inpainting\tnoteref{t1}}

\tnotetext[t1]{This work has received funding from the European Research 
	Council (ERC) under the European Union’s Horizon 2020 research and 
	innovation programme(grant agreement no. 741215, ERC Advanced Grant 
	INCOVID)}
 
\author[add1]{Rahul Mohideen Kaja Mohideen}\corref{cor1}
\ead{rakaja@mia.uni-saarland.de}

\author[add1]{Tobias Alt}
\ead{alt@mia.uni-saarland.de}

\author[add1]{Pascal Peter}
\ead{peter@mia.uni-saarland.de}

\author[add1]{Joachim Weickert}
\ead{weickert@mia.uni-saarland.de}

 \cortext[cor1]{Corresponding author}
\address{ Mathematical Image Analysis Group, Faculty of Mathematics 
and Computer Science\\
Campus E1.7, Saarland University, 
66041 Saarbr\"ucken, Germany \\
\{rakaja, alt, peter, weickert\}@mia.uni-saarland.de\\}

\begin{abstract}

Inpainting-based codecs store sparse selected pixel data 
and decode by reconstructing the discarded image parts by inpainting. 
Successful codecs (coders and decoders) traditionally use inpainting operators 
that solve 
partial differential equations. This requires some numerical expertise if
efficient implementations are necessary. Our goal is to investigate
variants of Shepard inpainting as simple alternatives for inpainting-based
compression. They can be implemented efficiently when we localise 
their weighting function. To turn them into viable codecs, we have
to introduce novel extensions of classical Shepard interpolation
that adapt successful ideas from previous codecs: Anisotropy allows 
direction-dependent inpainting, which improves reconstruction quality. 
Additionally, we incorporate data selection by subdivision as an 
efficient way to tailor the stored information to the image structure. 
On the encoding side, we introduce 
the novel concept of joint inpainting and prediction for isotropic Shepard
codecs, where storage cost can be reduced based on intermediate inpainting 
results. In an ablation study, we show the usefulness of these individual 
contributions and demonstrate that they offer synergies which elevate the 
performance of Shepard inpainting to surprising levels. Our resulting
approaches offer a more favourable trade-off between simplicity and 
quality than traditional inpainting-based codecs. Experiments show that 
they can outperform JPEG and JPEG2000 at high compression ratios.
\end{abstract}

\end{frontmatter}

\section{Introduction}

While inpainting was introduced to restore missing or damaged regions 
of an image~\cite{MM98a, EL99b, BSCB00}, inpainting-based 
compression~\cite{GWWB08, SPMEWB14} is unconventional in two aspects:
It drives inpainting to the extreme by considering very sparse data, and 
it combines it with 
data optimisation. In the encoding step, one stores only a small, 
carefully optimised fraction of the image pixels. During the decoding
phase, the unknown data are approximated by inpainting. Since inpainting-like
filling-in mechanisms are postulated to play an important role in the human 
visual system~\cite{We35}, inpainting-based compression appears natural
and conceptually appealing. 
Moreover, aiming at sparsity in the spatial domain is particularly
simple and distinguishes inpainting-based compression from widely-used 
transform-based approaches such as JPEG~\cite{PM92}, JPEG2000~\cite{CSE00}, 
and HEVC intra~\cite{Be14}. The latter ones aim at sparsity in the discrete 
cosine or wavelet domain. Advanced inpainting-based codecs can outperform
JPEG2000~\cite{SPMEWB14}, and they can be far ahead of the state-of-the-art
for data with a low to moderate amount of texture, such as depth 
maps~\cite{JPW21}.

\medskip
Apart from a few notable exceptions such as exemplar-based inpainting~\cite{KBPW18}, inpainting with concepts from Smoothed Particle Hydrodynamics~\cite{DAW21}, and linear spline inpainting~\cite{DNV97,DDI06,MMCB18}, most 
inpainting-based 
codecs (coders and decoders) employ partial differential equations (PDEs)
of diffusion type for inpainting. Homogeneous diffusion allows for very
efficient algorithms if one uses sophisticated numerical ideas~\cite{KSFR07, MBWF11, HPW15, CW21, KN22, KW22}, while edge-enhancing 
anisotropic diffusion 
offers highest quality due to its anisotropy~\cite{GWWB08,SPMEWB14}.

\medskip
Data optimisation has two aspects: One can optimise the positions of
the stored pixels {\em (spatial optimisation)} as well as their corresponding
grey or colour values {\em (tonal optimisation)}. The positions of the
stored pixels constitute the {\em inpainting mask}. While numerous
methods have been proposed for spatial and tonal optimisation, there
is a general trade-off between simplicity, efficiency, and quality.
 
\medskip
The goal of our paper is to set a new benchmark in inpainting-based
compression, that offers a better compromise than existing approaches
w.r.t.~simplicity of implementation, computational efficiency, and
approximation quality.
We aim at a family of simple and highly efficient codecs that does not 
require the numerical sophistication of PDE-based codecs, while keeping 
certain quality-critical features such as anisotropy and inpainting mask 
optimisation. It is based on variants of the classical Shepard 
interpolation idea~\cite{Sh68}.

\medskip
Shepard interpolation relies on the idea of normalised weighted averaging.
If one uses a localised weighting function, only a few surrounding mask 
pixels influence a given unknown pixel. The locality of the resulting Shepard 
inpainting allows simple and fast inpainting and tonal optimisation. This 
distinguishes it from PDE-based methods, where high efficiency in the
inpainting step is possible but requires the adaptation of advanced 
numerical concepts such as multigrid techniques~\cite{MBWF11, KSFR07}, Fast Explicit 
Diffusion~\cite{PSMMW15}, Green's functions~\cite{HPW15, KN22}, finite element 
methods~\cite{CW21}, and domain decomposition approaches~\cite{KW22}. 
Moreover, all exact methods for tonal optimisation of PDE-based approaches 
are relatively time-consuming and may be substantially slower than a
fast PDE-based inpainting step. Shepard interpolation is non-iterative 
and requires little numerical expertise. While the original paper by 
Shepard~\cite{Sh68} proposes inverse distance weighting functions without
localisation, we use a variant from~\cite{AAS17}, which employs a truncated
Gaussian weight function and only approximates the function values in the 
mask points. Since we perform tonal optimisation to maximise the 
approximation quality, this is unproblematic.


\subsection{Our Contribution}

Interestingly, despite its advantages, Shepard inpainting has not been
explored for compression prior to our conference publication~\cite{Pe19},
where we have shown that it allows highly efficient 
image compression with reasonable quality. Its usefulness has also been
confirmed for the compression of piecewise smooth images~\cite{JPW21}. 
In the present work, we extend and improve our results from~\cite{Pe19} by
fusing the best of both worlds: the high efficiency from Shepard inpainting 
with two quality improvements from successful PDE-based codecs 
\cite{GWWB08, SPMEWB14}, namely anisotropy and spatial mask adaptation. 
We also refine the work done in our conference publication by proposing 
a direct and more efficient tonal optimisation process for isotropic
Shepard inpainting.

\medskip
We propose a novel anisotropic version of Shepard inpainting which allows
elongated Gaussian kernels to adapt the inpainting direction to the local
image structure. Subdivision-based strategies enable us to find better
inpainting data than just regular masks, but are not as expensive to store
as fully optimised masks. For high compression ratios, our simple codec rivals 
the quality of transform-based approaches. In particular, it does not suffer 
from the pronounced block artefacts that plague the JPEG 
family~\cite{PM92,CSE00} at high compression rates.

\medskip
When being localised, our Shepard inpainting codecs can offer substantial 
speed-ups over most implementations
of PDE-based inpainting approaches. Our resulting methods are still 
simple and maintain a favourable trade-off between computational 
efficiency and reconstruction quality.


\subsection{Related Work}
\label{sec:related_work}

In this section, we discuss prior work regarding the three pillars of each 
successful inpainting-based compression pipeline: inpainting operators, 
data selection strategies, and encoding. 

\medskip
{\bf Inpainting Operators.} Since the inpainting operator recovers the 
missing image parts from the known data, it is crucial for the reconstruction 
quality. A significant number of operators use partial differential equations 
(PDEs). In
particular, homogeneous diffusion~\cite{Ii62} is a popular choice for 
compression~\cite{MBWF11,HMWP13,PHNH16,Ca88,GLC12,JPW20} since it is 
simple and fast compared to other PDE-based methods. It is an isotropic 
operator, i.e.~it propagates information from stored pixels equally in all 
directions. As a higher-order alternative to homogeneous diffusion, biharmonic 
inpainting~\cite{Du76} is another useful isotropic operator for sparse image 
inpainting~\cite{GWWB08,CRP14,SPMEWB14}. Last but not least, anisotropic 
variants of nonlinear PDEs have been explored, most notably edge-enhancing 
diffusion (EED)~\cite{We94e,WW06,GWWB08} and higher-order variants
\cite{PKW16,JS21a}. They adapt themselves to the local data structure. 
EED is the core component in some of the qualitatively best diffusion-based 
compression methods such as R-EED~\cite{SPMEWB14} and R-EED-LP~\cite{PKW16}. 

\medskip
While diffusion-based inpainting methods yield excellent results on 
piecewise smooth and mildly textured images, they struggle with 
high-frequent texture data. To address this issue, sparse exemplar-based 
inpainting methods have been proposed~\cite{FACS09}. They reconstruct 
images by copying pixels or whole 
image patches from similar neighbourhoods. It is also possible to combine 
diffusion and exemplar-based inpainting methods in hybrid codecs~\cite{PW15}.

\medskip
Deep learning-based inpainting methods have also generated 
interest~\cite{XXC12,PKDD16,YLLS17}. 
Successful concepts include generative adversarial networks (GANs)
~\cite{VHF21, Pe22} and deep priors approaches~\cite{UVL18}.
While deep learning approaches are undoubtedly powerful, 
they are computationally expensive to train and the models are not as 
transparent as PDE-based inpainting. 

\medskip
Shepard interpolation~\cite{Sh68} is a simple and straightforward 
inpainting operator. It can be interpreted 
as a special case of a class of inpainting operators known as radial basis 
functions (RBFs)~\cite{Ha71, Bu03} which have been successfully applied for 
scattered data interpolation~\cite{MD05, We05, AWA19, DFTT11}. Also, 
anisotropic variations of RBFs have been 
considered~\cite{BDL10,CMM10,DAW21}. 
The work of Daropoulos et al.~\cite{DAW21} is closest in spirit to our work, 
as they employ anisotropic RBF kernels with spatial and tonal optimisation. 

\medskip
Multiple publications have proposed improvements for isotropic Shepard 
interpolation. This includes restrictions of Shepard interpolation to a 
localised averaging of known data~\cite{FN80}, which is also crucial for our 
own applications. This influence area of known data has also been adapted 
locally~\cite{Re88,LZZZ+19}.

\medskip
In addition, strategies have been proposed to introduce directional information 
into Shepard interpolation. Tomczak~\cite{To98} assigns higher averaging 
weights to points that are aligned along locally dominant directions, whereas
Ringaby et al.~\cite{RFFO+14} use anisotropic distance
measures for image rectification. The approach of Lorenzi et al.~\cite{LSRM17} 
comes closest to our own anisotropic extension of Shepard interpolation. They 
define locally oriented ellipsoids in higher dimensions as averaging windows. 
However, compared to these previous methods, we benefit from the specific 
setting of our compression codec: instead of irregularly distributed known 
data, we define an anisotropic method on a regular grid. This allows more 
straightforward measures of anisotropy and a simple and elegant anisotropic 
extension of Shepard interpolation for the specialised purpose of compression.
Shepard interpolation has been used successfully in sparse image 
inpainting~\cite{KW93, AAS17}. However, our conference 
publication~\cite{Pe19} is the first that applies it to
compression.

\medskip
{\bf Data Selection.} While the inpainting operator plays a very significant 
role in the final reconstruction quality, choosing the right data is equally 
important.

\medskip
Spatial data optimisation can be broadly classified into two categories: 
unconstrained and constrained. Unconstrained mask selection approaches
~\cite{BBBW09,MHWT+11,ACV11,HSW13,CRP14,BLP+17,DCPC19,CW21,BHR21,Pe22,PSAW23} 
do not impose any structural 
restrictions on the mask pixel positions. This results in a mask which is 
completely optimised for the input image and the inpainting operator, 
which in turn results in the best possible reconstruction quality. 
However, from a coding viewpoint, this is not necessarily the best approach 
as storing these optimised positions can be very expensive. The compression 
of such unconstrained binary masks has been explored in detail in~\cite{MPW21}.

\medskip
On the other hand, constrained mask approaches place mask points such that 
the positions have some structure or predictability to them. The simplest 
strategy of which is to place points on a regular grid. Other options 
include hexagonal grids with unconstrained mask points~\cite{HMWP13}, and 
rectangular grids with edge information~\cite{JPW20,JPW21}. When dealing 
with regular masks in this work, we also consider rectangular grids.

\medskip
Subdivision-based approaches are another class of constrained mask 
approaches, where the mask is adapted to the image but the points are 
placed in a specific pattern. Usually, subdivision is implemented by 
splitting parts of the image with high error into smaller sub-images and 
placing more points where the reconstruction error is large. Earlier 
approaches such as~\cite{DNV97,DDI06,GWWB08} use a triangular subdivision, 
Later, Schmaltz et al.~\cite{SPMEWB14} proposed R-EED which employed 
rectangular subdivision, which was also used by R-EED-LP~\cite{PKW16}. 
Subdivision allows image adaptivity while offering efficient storage in the 
form of trees. This motivates us to consider subdivision for our Shepard 
inpainting-based compression pipelines. 

\medskip
Optimising the grey or colour values of the known pixels is called tonal 
optimisation~\cite{MHWT+11, HMHW+15}. The core principle of tonal optimisation 
is that errors are intentionally introduced in the known data such that the 
final reconstruction error is minimised. Related work on tonal optimisation
will be discussed in Subsection \ref{sec:tonal_opt}.

\medskip
{\bf Encoding.} After the mask locations and the corresponding pixel 
values have been selected, we need to compress this information to 
reduce the final file size further. Inpainting-based compression pipelines have 
used PAQ~\cite{Ma05} to great effect. PAQ is a family of entropy coders that 
employ context-mixing, wherein the symbol probabilities are calculated by 
combining probabilities from several similar symbol distributions that have 
been seen previously. In a way, we can interpret this procedure as trying to 
predict the next symbol which needs to be encoded. Prediction has been 
used in entropy coding to reduce file size~\cite{CW84}. 
It has also been integrated into the pipeline of popular image compression 
methods such as PNG~\cite{Cr95} and H.265-intra~\cite{Be14}. 

\medskip
However, for inpainting-based compression methods, prediction has only been 
performed as a part of the entropy coding, and it has never been integrated into 
the pipeline. Therefore, in this work, we exploit the nature of the inpainting 
operator to predict pixels during compression and decompression. This results 
in our joint inpainting and prediction approach.


\subsection{Organisation of the Paper}

In Section~\ref{sec:inp_compression}, we review the basic components of 
inpainting-based compression. In Section~\ref{sec:shepardanisotropic}, 
we introduce the anisotropic version of Shepard inpainting. After that, 
we present the full compression pipelines for our Shepard inpainting 
codecs in Section~\ref{sec:pipeline}. Finally, we evaluate the 
performance of these new approaches in Section~\ref{sec:experiments} and 
conclude our paper with a summary and an outlook on future work in 
Section~\ref{sec:conclusion}.


\section{Review: Inpainting-based Compression}
\label{sec:inp_compression}

Every inpainting-based compression approach requires a suitable inpainting 
method and a corresponding selection strategy for optimised known data. 
In this section, we review some of those approaches in more detail, and if 
they are either directly relevant to our own contributions. 
Moreover, we also discuss tonal optimisation as an additional way to 
optimise known data for better inpainting results. 


\subsection{Inpainting Methods}
\label{sec:inp_methods}

\subsubsection{Inpainting with Diffusion Processes}
Diffusion has a long tradition in image processing~\cite{Ii62, PM90, We97} 
and it has also been used for image inpainting; see e.g.~\cite{Ca88,WW06}. 
Let $f:\Omega \rightarrow \mathbb{R}$ denote a grey value image on a 
rectangular image domain $\Omega \subset \mathbb{R}^2$ that is only known 
on a subset $K\subset \Omega$, also called the {\em inpainting mask}. To 
reconstruct the unknown image data in $\Omega \setminus K$, diffusion-based
inpainting computes the steady state ($t \to \infty$) of the following 
initial value problem:
\begin{align}
 \partial_t u &= \textnormal{div}(\boldsymbol{D}\boldsymbol{\nabla} 
 {u}) &\textnormal{on}\  \, \Omega \setminus K \times (0,\infty ),  
\label{eq1}\\
{{u}}(x,y,t)&={f}(x,y,0) &\textnormal{on} \ \, K\times [0,\infty),  
\label{eq2}\\
\boldsymbol{n}^\top \boldsymbol{D}\boldsymbol{\nabla}{u}&=0 
&\textnormal{on}\  \, 
\partial\Omega \times(0,\infty) \, .  \label{eq3}
\end{align}
Here, $u(x,y,t)$ denotes the image pixel value at position $(x,y)$ and 
time $t$, and $\boldsymbol{n}$ is the outer normal vector at the image 
boundary $\partial \Omega$. The spatial gradient operator is denoted by 
$\boldsymbol{\nabla}=(\partial_x,\partial_y)^\top$, and  
$\textnormal{div}= \boldsymbol{\nabla}^\top$ is the divergence. 
The diffusion tensor $\boldsymbol{D} \in \R^{2\times2}$ is a positive 
semi-definite matrix. Its eigenvectors determine the propagation directions
of the diffusion process, and their eigenvalues determine the amount
of diffusion along those directions.
The Dirichlet boundary conditions in Eq.~\eqref{eq2} specify 
that known pixel values stay unmodified. Reflecting boundary conditions 
are defined in Eq.~\eqref{eq3} to avoid diffusion across the image 
boundaries.

\medskip
The simplest choice for the diffusion tensor is $\boldsymbol{D}=
\boldsymbol{I}$, where $\boldsymbol{I}$ is the identity matrix. In that 
case, we can write 
Eq.~\eqref{eq1} as
\begin{equation}
\partial_t {u} =\textnormal{div}(\boldsymbol{\nabla} {u}) = \Delta 
{u}= \partial_{xx}{u} + \partial_{yy}{u}.
\end{equation} 
This equation describes {\em homogeneous diffusion} which propagates  
information isotropically in all directions~\cite{Ii62}. 
	
\medskip
There are more sophisticated choices for $\boldsymbol{D}$ which also allow 
e.g.~direction-dependent (anisotropic) inpainting adaptation. For example, 
\emph{edge-enhancing anisotropic diffusion} 
(EED)~\cite{We94e, WW06} considers the diffusion tensor 
$\bm D(\bm \nabla u_\sigma)$, 
where $u_{\sigma}$ represents the convolution of the evolving image 
$u$ with a Gaussian kernel of standard deviation $\sigma$. This
Gaussian convolution makes the edge detector $|\bm \nabla u_\sigma|^2$
more robust under noise, where $|\,.\,|$ denotes the Euclidean norm. 
The first normalised eigenvector of $\bm D(\bm \nabla u_\sigma)$ is
chosen as $\bm v_1 = \bm \nabla u_\sigma / |\bm \nabla u_\sigma|$. 
It is perpendicular to the edge, while the second normalised eigenvector $\bm v_2$ 
is parallel to it. The eigenvalues $\mu_1$, $\mu_2$ denote the 
contrast in the direction of these eigenvectors. By setting $\mu_2=1$, 
one allows full diffusion along edges. To reduce diffusion across edges, 
one uses for $\mu_1$ a decreasing diffusivity such as the one by 
Charbonnier et al.~\cite{CBAB94}:
\begin{equation}
  \mu_1 = g\!\left(|\bm \nabla u_\sigma|^2\right) 
  = \frac{1}{\sqrt{1 + \frac{|\bm \nabla u_\sigma|^2}{\lambda^2}}}.
\end{equation}
with some contrast parameter $\lambda > 0$.
With these choices, $\bm{D}(\bm \nabla u_\sigma)$ can be written as
\begin{equation}
    \bm D(\bm \nabla u_\sigma) = g(|\bm \nabla u_\sigma|^2) 
    \bm v_1 \bm v_1^\top + \bm v_2 \bm v_2^\top.
\end{equation} 

\medskip
For inpainting with homogeneous diffusion or EED, one observes global
convergence where the steady state does not depend on the initialisation.
However, since a good initialisation can accelerate the convergence, a 
pragmatic approach is to initialise the non-mask pixels with the average 
grey value of the mask pixels.
 
\medskip
For EED inpainting, one discretises the parabolic PDE (\ref{eq1}) with 
finite differences and computes the reconstruction by means of numerical 
solvers~\cite{MM05}. An explicit time discretisation is simple but
has to obey severe time step size restrictions for stability reasons.
There are ways to accelerate explicit schemes by using cyclically 
varying time step sizes~\cite{WGSB16}, or extrapolation ideas~\cite{HOWR+16, TOW19}.
A semi-implicit time discretisation does not suffer from any time 
step size limits~\cite{We97}, but requires solving a large linear 
system with a matrix that is symmetric, positive definite, and sparse. 
To this end, one can use iterative solvers such as conjugate gradients.

\medskip
For homogeneous diffusion inpainting, efficient numerical solvers
often exploit direct discretisations of the Laplace equation $\Delta u = 0$
that arises in the steady state. 
This has been done with multigrid methods~\cite{MBWF11, KSFR07}, discrete
Green's function approaches~\cite{HPW15, KN22}, finite element discretisations 
with conjugate gradient solvers~\cite{CW21}, and domain decomposition 
algorithms~\cite{KW22}.
  
\medskip
These discussions show that diffusion-based inpainting requires quite 
some numerical expertise, if one aims at highly efficient algorithms. 
This motivates us to study alternatives that also offer efficient 
algorithms, but do not rely on such an expertise and 
lead to fairly simple implementations.

\subsubsection{Inpainting with Radial Basis Functions}

For inpainting with radial basis functions~\cite{Bu03, We05}, we consider the 
same interpolation problem as in the previous section, where
$f:\Omega \rightarrow \mathbb{R}$ where known data is only
given on the set $K \subset \Omega$. According to~\cite{Ha71}, this can be seen 
as a sparse interpolation 
problem that can be solved with a weighted averaging of the known data

\begin{equation}\label{eq:rbf}
    u(\bm x_i) = \sum_{\bm x_j \in K} w(\bm x_j - \bm x_i) c_j \, .
\end{equation}
Here, $w$ denotes a radial basis function, for instance a
multiquadric~\cite{We05}. The unknown coefficients $c_j$ of this interpolation 
approach are determined by the interpolation condition
\begin{equation}
    u(\bm x_j) = f(\bm x_j) \quad \forall \, \bm x_j \in K \, .
\end{equation}
Together with Eq.~\eqref{eq:rbf}, these constraints establish a linear system 
of equations that needs to be solved to obtain the coefficients $c_j$. As soon 
as these are known, the inpainting result is obtained directly by weighted 
averaging.

\subsubsection{Isotropic Shepard Inpainting}

Shepard interpolation~\cite{Sh68, KW93} can be interpreted as a simplified 
special case of normalised RBF interpolation. In particular, no interpolation 
weights need to be computed, and an inpainted value $u(\bm x_i)$ can be directly 
obtained as
\begin{equation}\label{eq:shep_orig}
     u(\bm x_i) =
    \begin{cases}
      \frac{\sum_{\bm x_j \in K} w(\bm x_j - \bm x_i) f_j}
      {\sum_{\bm x_j \in K} w(\bm x_j - \bm x_i)}, & \textnormal{if} \quad \forall 
      \  j: 
      |\bm x_i - \bm x_j| > 0 \, ,\\
      f_i, & \textnormal{if} \quad  \exists \  j: |\bm x_i - \bm x_j| = 0  \, .\\
    \end{cases}    
\end{equation}
The weighting function $w$ 
penalises the distance of the known data at $\bm x_j$ from the interpolated 
position $\bm x_i$ according to
\begin{equation}\label{eq:int_con}
    w(\bm x_j - \bm x_i) = \frac{1}{|\bm x_j - \bm x_i|^p} \, .
\end{equation}
The parameter $p>1$  controls 
the influence of neighbouring points, where higher values of $p$ result in 
less contributions from distant points.
Comparing Eq.~\eqref{eq:rbf} and 
Eq.~\eqref{eq:shep_orig}, the coefficients in Shepard interpolation are chosen 
as
\begin{equation}\label{eq:shepard_coeff}
  c_j:=\frac{f_j}{\sum_{\bm x_j \in K} w(\bm x_j - \bm x_i)} \, .
\end{equation}
For our compression application, a variant of the original definition of 
Shepard interpolation is helpful. This Shepard inpainting eliminates the case 
distinction in Eq.~\eqref{eq:shep_orig} and performs the weighted averaging 
from the first case everywhere~\cite{AAS17}. This relaxation implies that for 
known data, the interpolation condition is not necessarily met and thus our 
inpainting performs approximation instead of interpolation. In 
Sec.~\ref{sec:tonal_opt} we discuss why this choice is well-suited for our 
compression application. Moreover, in the following, we use Gaussian weights 
\begin{equation}\label{eq:gaussian}
  G_\sigma\left(\bm x\right):=\exp \left( \left( -|\bm{x}|^2)/(2\sigma^2 \right) \right) \, .
\end{equation}
Note that such a Gaussian weighting function has infinite support, and thus 
every known data point in the image would be used for each averaging. For our 
fast and simple codec, it is crucial to truncate these Gaussians. By limiting 
the weighting window to $(\lceil 4\sigma\rceil +1) \times (\lceil 4\sigma\rceil 
+1)$, the computational effort is reduced to $\mathcal(O)(\sigma^2 |K|)$. In 
addition, this truncation allows us to provide closed-form solutions for 
important optimisation steps in our encoder in Section~\ref{sec:rjip}.

\medskip

However, this truncation can lead to situations where the interpolation window 
does not contain any known data. To reduce this danger, we follow Achanta et 
al.~\cite{AAS17}, who adapt the standard deviation to 
the fraction of known data according to
\begin{equation} \label{eq:shepard_iso_sigma}
\sigma = \sqrt{(m \cdot n)/(\pi |K|)} \, .   
\end{equation}
Here $m$ and $n$ denote the dimensions of the discrete image, and $|K|$ is the 
number of mask pixels. This leads us to the final equation for Shepard 
inpainting of $u_i = u(\bm x_i)$:
\begin{equation}\label{eq:shepard_iso}
  u_i  = \frac{\sum_{\bm x_j \in K} G_\sigma(\bm x_j - \bm x_i) f_j}
  {\sum_{\bm x_j \in K} G_\sigma(\bm x_j - \bm x_i)}.
\end{equation}
Note that the weighting function only depends on the distance between the two 
pixels and not its orientation. Thus, we refer to this strategy as {\em isotropic} 
Shepard inpainting. In Section~\ref{sec:shepardanisotropic}, we extend 
this to an \emph{anisotropic} concept with oriented Gaussians which distribute 
information along dominant 
image structures.

 
\begin{figure*}[t]
    \centering

\begin{tabular}{c c c}

		\includegraphics[width = 0.3\textwidth]
		{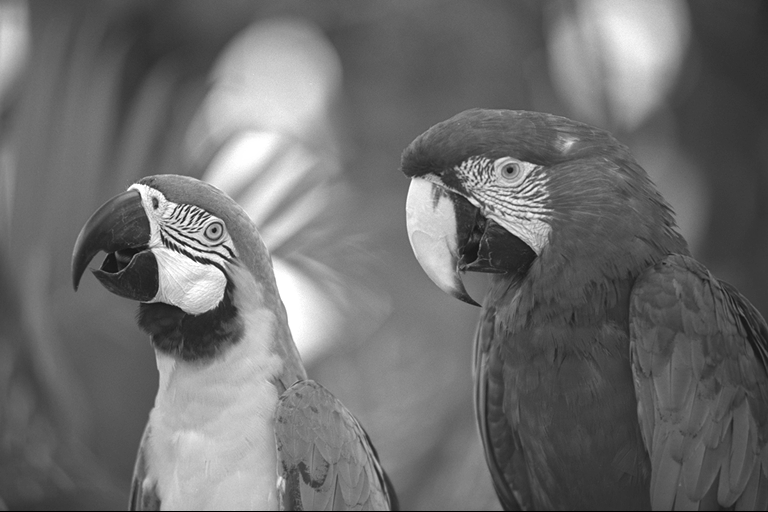} & 
		\includegraphics[width = 0.3\textwidth]
		{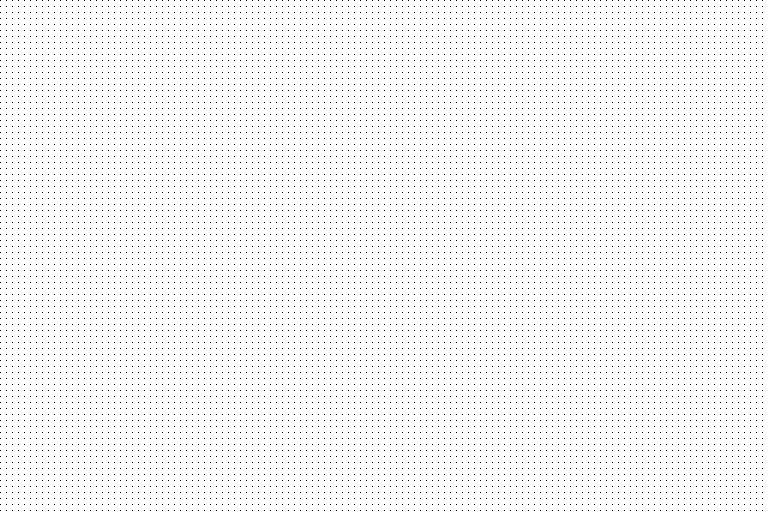} &
			\includegraphics[width = 0.3\textwidth]
		{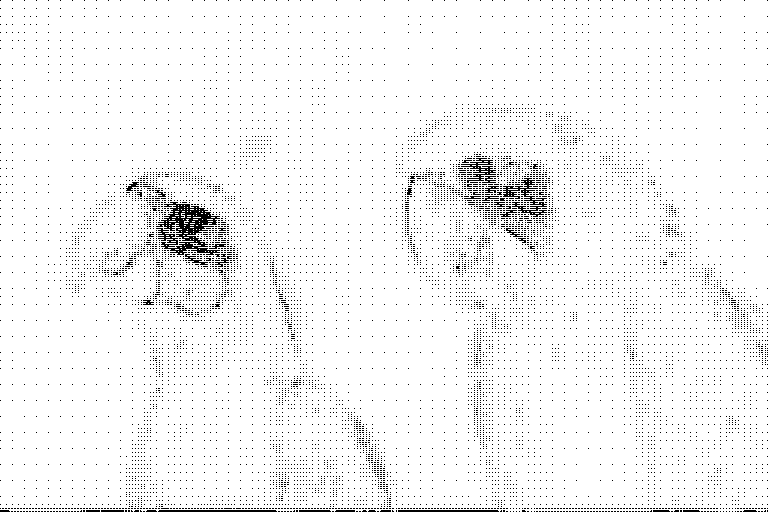} \\
		\emph{kodim23} & uniform mask & subdivision mask\\
        & for \emph{kodim23}&for \emph{kodim23}
		
\end{tabular}
\caption{Uniform masks do not adapt to the image structure but generate no 
overhead. On the contrary, subdivision can be used to create masks that are 
adaptive to 
the image structure and can be stored easily with trees.  
 The black pixels in the mask image are the known data points.}
\label{fig:subdivision}
\end{figure*}


 \subsection{Choosing Mask Points}
 
 \label{sec:inp_mask_op}

Spatial data selection has a large influence on the final reconstruction of 
an inpainting-based compression method. We can employ a variety of methods 
to select the known pixels.  Basic strategies rely on non-adaptive 
masks. For instance, they use a regular mask which places points on an uniform 
grid, or a (pseudo-)random mask. Both options require little to no storage 
cost, as we need to store only the grid size for regular masks or the seed 
value for the pseudo-random generator in the case of random masks. 

\medskip

There are also fully adaptive masks which do not have any constraints on mask 
 point positions but are very expensive to store~\cite{MPW21}. Finding such 
 adaptive masks constitutes a challenging optimisation problem which is 
 addressed by many diverse approaches~\cite{BBBW09, HSW13, BWD19, DB19, ACV11,
BLP+17,MHWT+11, APW17, CW21}. Recently, also deep learning has 
been explored to optimise locations of known data~\cite{DCPC19, PSAW23}. 

\medskip

Subdivision masks~\cite{SPMEWB14, PKW16, DNV97} offer a compromise between the 
two extremes mentioned above. They offer a good trade-off 
between spatial adaptivity and ease of compression. Subdividing the image and 
storing points at fixed positions in these subimages reduces the storage cost 
significantly. An example subdivision mask can be seen in 
Fig.~\ref{fig:subdivision}.

\medskip

Since we aim for simple and fast codecs, we use only non-adaptive and 
subdivision masks in the following.

\subsection{Tonal Optimisation}
\label{sec:tonal_opt}

In addition to choosing the location of mask pixels, it is also 
possible to optimise their value. This is referred to as tonal 
optimisation~\cite{MHWT+11, HMHW+15, Ho16}. It modifies the known pixel values 
such that the global mean squared error (MSE) is minimised. Even though this 
introduces errors in the sparse stored data, this is outweighed by significant 
reconstruction improvements in the large unknown areas. 

\medskip

Since the known data is no longer reliable, it makes sense to relax the 
interpolation condition. Therefore, 
we choose approximation over interpolation for our inpainting. Previous 
approaches have addressed this issue with post-processing instead 
\cite{SPMEWB14}. 
In addition to tonal optimisation, there are more sources for errors in known 
data: noise and quantisation. Coarse 
quantisation intentionally reduces the number of admissible pixel values in 
the grey-level domain, hence reducing storage costs.   
Quantisation can be applied as post-processing after tonal optimisation 
but might revert some of its improvements. Thus, it is often preferred to 
account for quantisation already during the tonal optimisation. 

\medskip

For linear inpainting operators such as homogeneous diffusion, we can write 
the tonal optimisation problem as an energy minimisation problem, and there 
are various strategies to solve it. A detailed review of these strategies can 
be found in~\cite{PSAW23}.

\medskip

A simple but effective tonal optimisation method that can be applied to all 
deterministic inpainting operators relies on the principle of trial and error. 
It visits known pixels in a random order and adjusts their quantised values to 
a higher or lower level. Changes that yield a lower inpainting 
error~\cite{SPMEWB14} are kept. This method is widely applicable and takes 
quantisation into account directly, but also often entails a high cost since 
every quality check requires an image inpainting. In Section~\ref{sec:rjip},
we show that in combination with Shepard inpainting, this method can be highly 
effective. 

\section{Anisotropic Shepard Inpainting}

\label{sec:shepardanisotropic}

Shepard inpainting uses isotropic Gaussian functions which 
are rotationally invariant. Thus, information from known pixels is propagated 
equally in all directions. However, additional directional 
information is implicitly encoded in the known pixels, and the use of such 
data has been successful in anisotropic diffusion inpainting~\cite{WW06,
GWWB08}. This motivates us to combine the speed and simplicity of Shepard 
inpainting with ideas from edge enhancing diffusion (EED)~\cite{We94e}.

\medskip

To this end, we introduce \emph{anisotropic Shepard inpainting}, which allows 
the weighting function to adapt to the local directional structure of the 
available data. We compute gradient information from the available data and, 
in the same fashion as EED, use this information to guide the influence 
function accordingly. Thus, we achieve an anisotropic inpainting effect. In 
particular, when the known data are arranged in a regular fashion, we can 
compute the gradient information without any overhead.  

\medskip

We propose to modify the weighting function $w$ based on structural 
information which is encoded in the vector containing the mask pixels $\bm f \in \mathbb{R}^{mn}$ for an image of resolution $m \times n$. To this end, we adapt the weighting 
function at each mask position $\bm x_j \in K$ to obtain a set of functions 
$w_j$. Our anisotropic Shepard inpainting computes the reconstruction 
$u_i$ as 
\begin{equation}\label{eq:Shepard_aniso}
  u_i = \frac{\sum_{\bm x_j \in K} w_j(\bm x_j - \bm x_i) f_j}
  {\sum_{\bm x_j \in K} w_j(\bm x_j - \bm x_i)}.
\end{equation}
The spatially varying weighting functions $w_j$ depend on the local structure 
of the masked image $\bm f$. 

\medskip

For simplicity, we now switch to the continuous case and consider a 
greyscale image $f: \Omega \rightarrow \mathbb{R}$ which is fully available 
on a continuous domain $\Omega\subset\mathbb{R}^2$. The gradient 
$\bm \nabla f$ encodes the structural information of $f$ which allows to 
define structure-adaptive weighting functions. 

\medskip

As a weighting function we choose an oriented Gaussian with 
standard deviations $\sigma_1, \sigma_2$, and a rotation angle $\theta$. The 
two standard deviations $\sigma_1, \sigma_2$ determine the major and minor 
directions of the Gaussian. For $\sigma_1 = \sigma_2$, we want to obtain a 
rotationally invariant Gaussian corresponding to the isotropic case. 

\medskip

\begin{figure}[t]
  \centering
  \hspace{1.5cm}
  \begin{tikzpicture}[scale=1.7]
  \clip (1.0, 1.2) rectangle (6,5.5);
  \draw [black, name path=one] plot [smooth, tension=1] coordinates 
  {(0,2.4) (3.4,3.4) (5,6)};
  
   \draw [draw=none, name path=two] plot [smooth, tension=0.01] 
   coordinates 
   {(0,2.4) (0,6) (5,6)};
   \tikzfillbetween[
     of=one and two,split, on layer=main
   ] {fill,opacity = 0.05};
    
   \coordinate (p1) at (3.4, 3.36);
   \coordinate (p2) at (4.6, 4.3);
   \coordinate (p3) at (4.34,2.16);
    
   \draw[thick,-latex] (p1) -> (p2);
   \draw[thick,-latex] (p1) -> (p3);
   \node[yshift=-3mm]  at (p3)  {$\bm \nabla f$};
   \node[xshift=4mm]  at (p2)  {$\bm \nabla^\bot f$};
      
   \coordinate[right of = p1, xshift=5mm] (x); 
   \coordinate[below of = p1, yshift=-5mm] (y); 
      
   \draw[dashed,-latex] (p1) -> (x);
   \draw[dashed,-latex] (p1) -> (y);
   \node[yshift=-2mm]  at (y)  {$x$};
   \node[xshift=2mm]  at (x)  {$y$};
   \draw pic[draw=black, ->, angle eccentricity=0.8, angle 
   radius=1.1cm]
        {angle=x--p1--p2};
   \node[xshift=8mm,yshift=3mm] at (p1) {$\theta$};

   \draw[rotate around={-54:(p1)},blue] (p1) ellipse (0.7cm 
   and 1.55cm);
   \draw[blue, |-|] (1.7,3.2) -- node[above] {$\sigma_1$} 
   (4.2,5.0);
   \draw[blue, |-|] (1.6,2.9) -- node[below,xshift=-2mm] 
   {$\sigma_2$} (2.4,1.8);
  \end{tikzpicture}
  \caption[Local adaptation of the weighting function to the image 
  structure]{Local adaptation of the weighting function to the image 
  structure in a continuous setting. The gradient $\bm \nabla f$ spans 
  a coordinate system rotated by $\theta$ w.r.t. the $(x,y)$-system. 
  The gradient magnitude shrinks the level lines of the Gaussian kernel 
  (blue) across dominant structures. This gives elliptic level lines 
  with major and minor axes proportional to $\sigma_1$ and $\sigma_2$, 
  respectively.
  \label{fig:shepaniso:shepard_system}}
\end{figure}
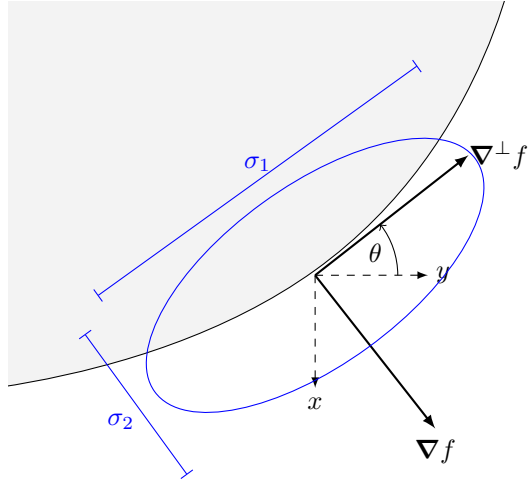

A model which fulfils the above properties based on the structural 
information described by the image gradient 
$\bm \nabla f = (f_{x}, f_{y})^\top$ 
in the $(x,y)$-coordinate system is given by
\begin{align}
  \label{eq:sig1}
  \sigma_1^2 &= \sigma^2, \\
  \label{eq:sig2}
  \sigma_2^2 &= g\!\left(|\bm \nabla f|^2\right) \sigma^2, \\
  \label{eq:angle}
  \theta &= \arctan\left(-\frac{f_{x}}{f_{y}}\right),
\end{align}
where $\sigma$ is an input parameter determining the base standard deviation 
of the Gaussian as in the isotropic case. Note that $\sigma_1$ and $\sigma_2$ 
are functions of $\sigma$ and $\bm \nabla f$, but have been abbreviated for 
readability. In Fig.~\ref{fig:shepaniso:shepard_system}, we visualise the 
relation between these parameters in a continuous setting.

\medskip

In our experiments, we found the rational Perona--Malik 
diffusivity~\cite{PM90}

\begin{equation} \label{eq:perona_malik}
  g\!\left(s^2\right) = \frac{1}{1 + \frac{s^2}{\lambda^2}}
\end{equation}
 to be a suitable choice. It attenuates the variance $\sigma^2$ at 
 image locations with dominant structures where the edge detector 
 $|\bm \nabla f|$ exceeds some contrast parameter $\lambda$. Choosing 
 the constant diffusivity~\cite{Ii62} $g(s^2)=1$ returns the isotropic 
 Shepard inpainting model. 

\medskip

The angle $\theta$ determines the rotation of the anisotropic Gaussian 
function. As $\bm \nabla f$ points into the direction of the steepest 
ascent of $f$, the anisotropic Gaussian should be oriented along the 
orthogonal direction $\bm \nabla^\bot f$. In 2D, a vector which is 
orthogonal to the gradient can be easily constructed, e.g. as 
$\bm \nabla^\bot f = (f_{x}, -f_{y})^\top$. The angle of this vector 
in the respective coordinate system is given by $\theta$ in 
Eq.~\eqref{eq:angle}.

\medskip

In the $(\bm\nabla^\bot f, \bm\nabla f)$-coordinate system, the 
resulting Gaussian weighting function should scale the kernel along the 
principal directions by the standard deviations $\sigma_1,\sigma_2$ given 
above. Thus, we obtain
\begin{equation}
\bm z^\top \bm \Sigma \bm z =
\begin{pmatrix}
z_1 & z_2
\end{pmatrix}
\begin{pmatrix}
\frac{1}{2\sigma_1^2} & 0 \\ 0 & \frac{1}{2\sigma_2^2}
\end{pmatrix}
\begin{pmatrix}
z_1 \\ z_2
\end{pmatrix}
\end{equation}
as an argument for the Gaussian function, where $\bm z$ is the spatial
difference between to positions in the $(\bm\nabla^\bot f, \bm\nabla
f)$-coordinate system.

\medskip

Let us now bring this argument to the $(x,y)$-coordinate system. To this 
end,we multiply a rotation by $\theta$ from the right, and its inverse 
from the left, yielding
\begin{equation}
\bm R_\theta^{-1} \bm z^\top \bm \Sigma \bm z \bm 
R_\theta = \bm d^\top \bm
R_\theta^{-1} \bm \Sigma \bm R_\theta \bm d.
\end{equation}
Here, $\bm d$ is the spatial distance between two positions in
the $(x,y)$-coordinate system. Note that we used the relation 
$\bm z = \bm R_\theta \bm d$ to move the rotation matrices inside the 
expression.

\medskip

Expressing the inner matrix with parameters $\alpha,\beta,\gamma$, we 
obtain

\begin{align}
&\begin{pmatrix}
\alpha & \beta \\ \beta & \gamma
\end{pmatrix}
=\bm R_\theta^{-1} \bm \Sigma \bm R_\theta
\end{align}
which, by additionally using the identity 
$\cos(\theta)\sin(\theta) = \frac12\sin(2\theta)$ yields
\begin{align}
\alpha(\theta, \sigma_1, \sigma_2) &= \frac{\cos^2(\theta)}{2 \sigma_1^2}
+\frac{\sin^2(\theta)}{2 \sigma_2^2}, \\
\beta(\theta, \sigma_1, \sigma_2) &= -\frac{\sin(2\theta)}{4 \sigma_1^2}
+\frac{\sin(2\theta)}{4 \sigma_2^2}, \\
\gamma(\theta, \sigma_1, \sigma_2) &= \frac{\sin^2(\theta)}{2 \sigma_1^2}
+\frac{\cos^2(\theta)}{2 \sigma_2^2}.
\end{align}
The resulting Gaussian weighting function takes the distance between two
spatial positions $\bm d = \left(d_1, d_2\right)^\top= \bm x - \bm y$ and
computes
\begin{equation}
G_{\theta,\sigma_1,\sigma_2}(\bm d) = \exp\!\left(- \alpha d_x^2 + 2 \beta 
d_x d_y - \gamma d_y^2\right),
\end{equation}
where $\alpha$, $\beta$ and $\gamma$ are functions of $\theta$, $\sigma_1$ 
and $\sigma_2$.

\section{Compression Pipeline}

\label{sec:pipeline}

\subsection{Regular Grid Codec with Isotropic Shepard inpainting}
\label{sec:rjip}

As a first method that aims for maximal simplicity, we propose the 
\emph{regular grid codec with joint inpainting and prediction} (RJIP), 
which we presented in our conference publication~\cite{Pe19}. While it 
is not image adaptive, it exploits novel prediction principles with 
inpainting instead. We store the known data on a regular mask, which 
means the only storage cost for positional data is the grid size 
parameter $r$. This generates minimal overhead. For the grey value data 
corresponding to the mask positions, we use an equally straightforward 
uniform scalar quantisation: We map the 8 bit colour values to a reduced 
range range $\{0,\ldots,q-1\}$ by partitioning the tonal domain into $q$ 
subintervals of equal length. 

\medskip

Shepard inpainting from Eq.~\eqref{eq:shepard_iso} is implemented by 
visiting each mask point $\bm x_j \in K$ and adding its impact on the 
reconstruction. The contribution to the numerator is stored in the value 
accumulation map $\bm v$ and the contribution to the denominator is stored 
in the weight accumulation map $\bm w$. For $w_i:=w(\bm x_i)$ and 
$v_i:=v(\bm x_i)$, both maps are then updated as follows: 

\begin{equation}
    w_i \leftarrow w_i + G_\sigma(\bm x_i - \bm x_j),
\end{equation} and 
\begin{equation}
    v_i \leftarrow v_i + G_\sigma(\bm x_i - \bm x_j)f_j
\end{equation} 
for all points $\bm x_i$ in the truncated Gaussian neighbourhood 
$\mathcal{N}_j$ of $\bm x_j \in K$. The new inpainting at point $\bm x_i$ 
can then be computed as 
\begin{equation}
    u_i = v_i/w_i. 
\end{equation}

To further decrease the final compressed file size, RJIP employs 
\emph{joint inpainting and prediction}.
In general, often better coding efficiency can be achieved by predicting the 
future values to be stored from already encoded data. Instead of the values 
themselves, we can encode the prediction error, the so-called \emph{residuals}, 
instead. 
Good predictions yield residuals that cluster around zero, thus reducing 
entropy, which directly translates to reduced storage cost. 

\medskip

We integrate this idea seamlessly into Shepard inpainting.
During image compression, the mask points are traversed one-by-one.
If the weight accumulation map $w$ is non-zero at the location the next
mask point to be encoded, an initial prediction can computed through 

\begin{equation}
    p_i = v_i/w_i \, .
\end{equation} 
Then we encode the residual between the prediction and the actual mask 
value as \begin{equation}
    e_i = (p_i - f_i)\, \textnormal{mod}\, q \, .
\end{equation}

Using a sufficiently large Gaussian as in Eq.~\eqref{eq:shepard_iso_sigma} 
ensures that each new encoded data point can predict at least one data point 
that has not yet been encoded. After the mask point
is encoded, we update the value and accumulation maps and repeat until all 
points are encoded. We finally compress the residuals with a suitable entropy 
coder. RJIP relies on \emph{finite state entropy} (FSE)~\cite{Co14}, a fast 
coder similar to arithmetic coding~\cite{WNC87}.

\medskip

For a given compression ratio, RJIP chooses the parameters $r$ and $q$ through 
a golden-section search such that the best reconstruction quality for the 
desired file size is obtained. While the iterative random walk method that we 
discussed in Section~\ref{sec:tonal_opt} is already well-suited for Shepard 
inpainting due to its locality, it even allows a closed-form solution for the 
tonal optimisation of individual pixels. 

\medskip

In the following,
 $u^{\textnormal{old}}_i$ and $u^{\textnormal{new}}_i$ are the old and new 
 pixel value at $\bm x_i \in K$. We want to find $u_i^\text{new}$, such that 
 it minimises the \emph{mean squared error} (MSE). Then the new error after 
 updating $u^{\textnormal{old}}_i$ to $u^{\textnormal{new}}_i$ is given by
 ~\cite{MPAWS20}

\begin{equation}\label{eq:rjip_error}
e (u_i^\text{new})  = \sum_{\bm x_j \in \mathcal{N}_i }
\left(f_j - \frac{v_j + G_\sigma(\bm x_j - \bm x_i) \left(u_i^\text{new} - 
	u_i^\text{old}\right)}{w_j}\right)^2,
\end{equation}
where $\mathcal{N}_i$ is the neighbourhood of points around 
 $\bm x_i$, $v_j$ and $w_j$ are the value accumulation map and the weight 
 accumulation map from Eq.~\eqref{eq:shepard_iso}, $G_\sigma$ is the Gaussian 
 defined in Eq.~\eqref{eq:gaussian}, and $f_j$ is the ground truth value at 
 $\bm x_j$. Intuitively, the error can be computed by subtracting the 
 weighted old mask value in the neighbourhood and adding the weighted new 
 mask value.

\medskip

An optimal tonal value should minimise this error. As the error function 
described in Eq.~\eqref{eq:rjip_error} is convex, an optimal tonal value is 
obtained by the condition $\frac{d}{d u_i^\text{new}} e(u_i^\text{new})=0$. 
This yields the closed-form solution

\begin{equation}
u_i^\text{new} = \frac{\sum_{\bm x_j \in \mathcal{N}_i } 
	\frac{G_\sigma(\bm x_j - \bm x_i)}{w_j}
	\left(f_j - \frac{v_j - G_\sigma(\bm x_j - \bm x_i) u_i^\text{old}}
	{w_j}\right)}
{\sum_{\bm x_j \in \mathcal{N}_i}
	\frac{G_\sigma(\bm x_j - \bm x_i)^2}{w_j^2}} \, .
\end{equation}

It enables us to directly compute the optimal tonal value at a mask point. 
Then 
we project the obtained optimal values to the set of quantised values. 
Iterating these computations over all mask points multiple times converges 
to optimised mask values. As we do not have to compute the inpainting for 
the full image every time we change a pixel value, the tonal optimisation 
for isotropic Shepard inpainting is highly efficient.

\subsection{Subdivision Codec with Anisotropic Shepard inpainting}

\label{sec:subdivision}
Subdivision masks offer a good balance between image adaptivity, coding 
costs, and complexity, which make them an attractive component of our 
codecs. We adopt this concept from Schmaltz et al.~\cite{SPMEWB14} by 
starting with the whole image as a single block and placing a mask point 
in each corner of the image. If the reconstruction error in a block is 
higher than a threshold parameter, we split the block in two along its 
largest dimension and add mask points at each corner of the smaller 
blocks. This process is then repeated until all blocks have a 
reconstruction error lower than the given threshold. We can then store 
the splitting decisions in the form of a binary tree which has a low 
coding cost. 

\medskip

To overcome the issue of approximating derivatives on a non-uniform 
grids, we first perform an isotropic Shepard inpainting on the mask and 
then compute derivatives with a sampling distance of one on the inpainted 
image. Then we use these derivatives to compute the final anisotropic 
inpainting. 

\medskip

Uniform mask codecs use a global Gaussian standard deviation $\sigma$ in 
Eq.~\eqref{eq:shepard_iso_sigma}. However, if we use a global $\sigma$ 
that ensures no holes in sparse image regions, regions with high mask 
density tend to have overly smooth reconstructions.  Therefore, in the 
case of subdivision mask codecs, we adapt $\sigma$  to the local 
density of the mask. Unfortunately, storing individual variance values 
for each mask point explicitly would be too costly. Instead, we derive 
the local variance from the mask itself following~\cite{WHL10}:

\begin{equation}
    \sigma_k = (\log(1 + A_k))^p \, .
\end{equation} 

Here, $\sigma_k$ is the variance at mask point $k$ and $p$ is a constant. 
To obtain $A_k$, we first perform a Voronoi decomposition~\cite{AKL13} of 
the image with the mask points as cell centres. Then we compute $A_k$ as 
the area of the Voronoi cell, which is the collection of points around a 
mask point $k$ that are closer to it than any other mask point. Therefore, 
instead of optimising and storing individual values for $\sigma$, we optimise and store 
$p$ instead. 

\medskip

We also have to optimise for the contrast parameter $\lambda$ for the 
diffusivity function. As $\lambda$ only determines the degree of anisotropy 
for all kernels, it does not need to be adapted locally.  

\medskip

Unlike RJIP, where we had a target ratio as a model parameter, we optimise 
our subdivision and quantisation w.r.t.~a target splitting error. We first 
find the quantisation parameter $q$. To that end, we consider the curve that 
maps quantisation levels to the corresponding quantisation error for the 
original image. As the number of quantisation levels increases, the quantised 
image gets closer to the original image and the quantisation error decreases, 
which implies that the quantisation levels vs.~error curve is decreasing. We 
select the value of $q$ such that the curve has a derivative value of $1$. 
This value of $q$ is the point of diminishing return, after which increasing 
$q$ does not result in a significant reduction in quantisation error. This 
ensures that $q$ is large enough to have a reasonable quantisation error but 
not so large to have a large coding cost.

\medskip

After fixing $q$, we perform the actual subdivision. We start at the root 
of the tree and reconstruct the image with just the four corner points of 
the image with anisotropic Shepard inpainting with optimised $p$ and 
$\lambda$ and then tonal optimisation. The parameter optimisation and tonal optimisation 
are alternated to adapt the parameters and the tonal values to each other so 
that the inpainting quality is increased. If the reconstruction error is 
higher than the target splitting error, we split the node and go to the 
next level. This process of adjusting the parameters $p$ and $\lambda$, the tonal 
optimisation, and node splitting is repeated for each tree level by going 
deeper into the tree until all sub-images or leaf nodes of the tree have a 
reconstruction error lesser than the target splitting error. Finally, the 
subdivision tree and the mask values are compressed by applying LPAQ2
~\cite{lpaq}. An overview of the algorithm can be seen in 
Algorithm~\ref{alg:subdivision}.

\begin{algorithm}[t]
\begin{algorithmic}[1]
\State compute $q$ from the quantisation error curve
\While{number of nodes split $>$ 0}
\State place mask points at leaf nodes
\While{$n < \textnormal{iter\_max}$, $\textnormal{iter\_max} \in 
\mathbb{N}$}
\State optimise for $\lambda$ and $p$
\State perform tonal optimisation
\EndWhile
\State compute reconstruction error at each leaf node
\State split leaf nodes with error $>$ target splitting error
\EndWhile
\State compress tree and mask values with LPAQ
\end{algorithmic}
\caption{Summary of the subdivision codec with anisotropic Shepard 
inpainting}
\label{alg:subdivision}
\end{algorithm}

\section{Experiments}
\label{sec:experiments}

In this section, we perform a systematic evaluation of 
our inpainting operators and compression pipelines in four parts. First, 
we compare our proposed operator to existing inpainting operators 
on a synthetic disc image both w.r.t. inpainting quality and runtime. Afterwards, we evaluate the scaling behaviour of the computational cost
w.r.t.~the number of pixels on an image from \emph{Sintel}~\cite{Ro11}. The third part is dedicated to a comparison of our uniform mask-based codecs 
to transform-based codecs on the test image \emph{trui} and the greyscale 
version of the Berkeley dataset~\cite{MFTM01}. Finally, we evaluate our subdivision-based 
codecs on the greyscale version of the Kodak dataset~\cite{Ko99} in comparison 
to JPEG.

\subsection{Comparing Inpainting Operators}
\label{sec:inp_comp}

\begin{figure}
    \centering

    \begin{tabular}{c c}
    \multicolumn{2}{c}{Input data} \\
       \begin{tikzpicture}[spy using outlines={red,magnification=3,size=1cm}]
		
\node {\includegraphics[width = 0.21\textwidth]
{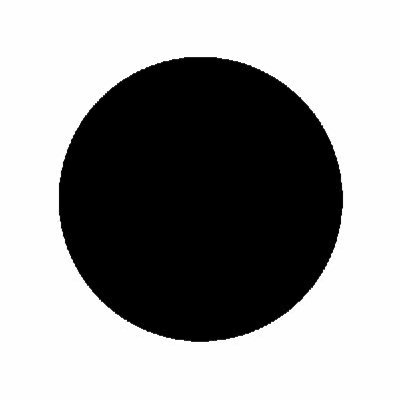}};
\spy on (0.6,-0.65) in node [left] at (2,-1);
\end{tikzpicture}
& 
 \begin{tikzpicture}[spy using outlines={red,magnification=3,size=1cm}]
		
\node {\includegraphics[width = 0.21\textwidth]
{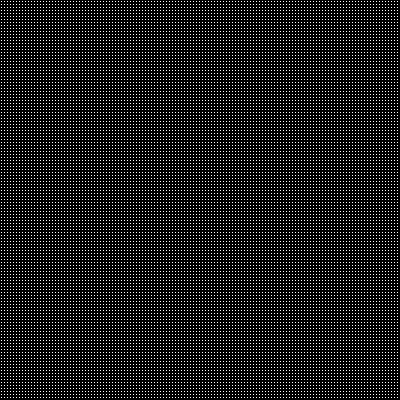}};
\spy on (0.6,-0.65) in node [left] at (2,-1);
\end{tikzpicture}
\\

\emph{disk} & mask  \\
\hline \\[0.1cm]

    \end{tabular}
    
\begin{tabular}{c c c}

\multicolumn{3}{c}{Homogeneous diffusion} \\
\begin{tikzpicture}[spy using outlines={red,magnification=3,size=1cm}]
		
\node {\includegraphics[width = 0.21\textwidth]
{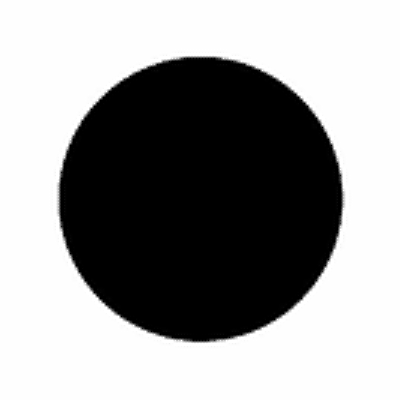}};
\spy on (0.6,-0.65) in node [left] at (2,-1);
\end{tikzpicture} &

\begin{tikzpicture}[spy using outlines={red,magnification=3,size=1cm}]
		
\node {\includegraphics[width = 0.21\textwidth]
{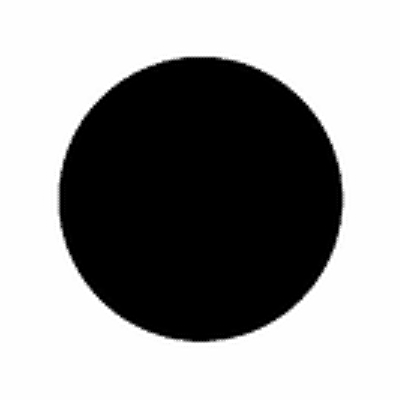}};
\spy on (0.6,-0.65) in node [left] at (2,-1);
\end{tikzpicture} &

\begin{tikzpicture}[spy using outlines={red,magnification=3,size=1cm}]
		
\node {\includegraphics[width = 0.21\textwidth]
{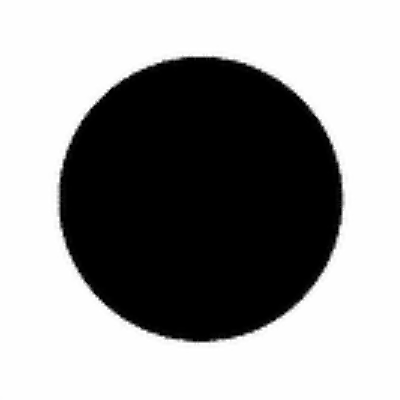}};
\spy on (0.6,-0.65) in node [left] at (2,-1);
\end{tikzpicture}  
\\ 
Greens functions & finite difference method & finite element \\ 
  with a Cholesky solver & with a CGNR solver & method\\
TO $\approx$ 1113 s & TO $\approx$ 0.32 s & TO $\approx$ 0.15 s\\
 inpainting $\approx$ 7.9 s & inpainting $\approx$ 0.03 s & inpainting $\approx$ 0.04 s\\
 MSE = 65.57 & MSE = 65.61 & MSE = 80.37 \\
 \hline \\[0.1cm]

\end{tabular}

\begin{tabular}{c c}
\multicolumn{2}{c}{Shepard inpainting} \\

\begin{tikzpicture}[spy using outlines={red,magnification=3,size=1cm}]
		
\node {\includegraphics[width = 0.21\textwidth]
{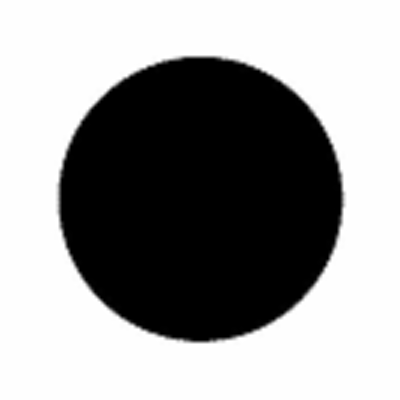}};
\spy on (0.6,-0.65) in node [left] at (2,-1);
\end{tikzpicture} &

\begin{tikzpicture}[spy using outlines={red,magnification=3,size=1cm}]
		
\node {\includegraphics[width = 0.21\textwidth]
{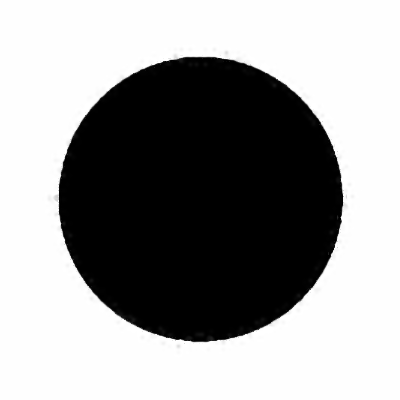}};
\spy on (0.6,-0.65) in node [left] at (2,-1);
\end{tikzpicture} 
		
\\ 
isotropic & anisotropic \\
TO $\approx$ \textbf{0.036 s} & 
TO $\approx$ 54.69 s\\
inpainting $\approx$ \textbf{0.004 s} & inpainting $\approx$ 0.024 s \\
 MSE = 92.85 & MSE = \textbf{16.35}
		
\end{tabular}
\caption{We compare the inpainting quality, time for tonal optimisation (TO), and inpainting time of Shepard inpainting to several contemporary solvers for 
homogeneous diffusion inpainting on a synthetic image of a disk of size $400 
\times 400$. All approaches use the same uniform mask of density $11.11\%$ (1 in 
9 pixels). The experiments were run on a single core of a single core of an 
Intel Core i7-6700A@3.40GHz with 32 GB RAM. The experiment highlights sharper 
results of anisotropic Shepard inpainting compared to all competitors at an 
inpainting speed comparable to homogeneous diffusion. Localised isotropic Shepard 
inpainting can be up to an order of magnitude faster than non-localised homogeneous 
diffusion inpainting at a similar quality. }
\label{fig:inpainting}
\end{figure}

In Fig.~\ref{fig:inpainting}, we consider the inpainting quality and 
computational effort on a synthetic binary disk image with a uniform mask of 
density $11.11\%$. In addition to our isotropic 
Shepard and anisotropic Shepard inpainting, we also consider homogeneous 
diffusion inpainting~\cite{Ii62} as a baseline since it is widely 
used in compression applications due to its relative simplicity compared to 
more sophisticated diffusion inpainting. Thus, it is the direct competitor for 
our Shepard inpainting. To ensure a fair comparison w.r.t. timing, we use 
several contemporary solvers for homogeneous diffusion~\cite{Ho16,CW21}.

\medskip

From Fig.~\ref{fig:inpainting}, we can see that the isotropic Shepard 
inpainting result yields slightly blurrier edges than homogeneous diffusion. 
However, the inpainting is faster by one to three orders of magnitude. Tonal 
optimisation can even be faster by up to five orders of magnitude compared to 
the approach of Hoffmann~\cite{Ho16}. Even the recent highly efficient finite 
elements method is still slower by one order of magnitude. This is a direct 
effect of the closed-form solution from Section~\ref{sec:tonal_opt} for 
isotropic Shepard inpainting. Thus, it constitutes a good alternative to 
homogeneous diffusion inpainting for time-critical applications while being 
significantly easier to implement.

\medskip

The anisotropic version of Shepard inpainting yields much sharper results than 
its competitors. However, this comes at the price of a higher computational load. 
For pure inpainting, anisotropic Shepard is still faster than homogeneous 
diffusion by one to three orders of magnitude depending on the solver. However, 
homogeneous diffusion can be faster for tonal optimisation. Here, the 
anisotropy prevents the efficient closed-form solution and requires a fall-back 
to a simple trial-and-error algorithm. Thus, anisotropic Shepard is a good 
choice for pure inpainting applications and compression which requires high 
quality.

\subsection{Timing Experiments}

In this experiment, we investigate the scaling behaviour of our Shepard 
inpainting-based compression methods and homogeneous diffusion with 
the number of pixels. We consider RJIP and a uniform mask codec with 
anisotropic Shepard inpainting.

\medskip

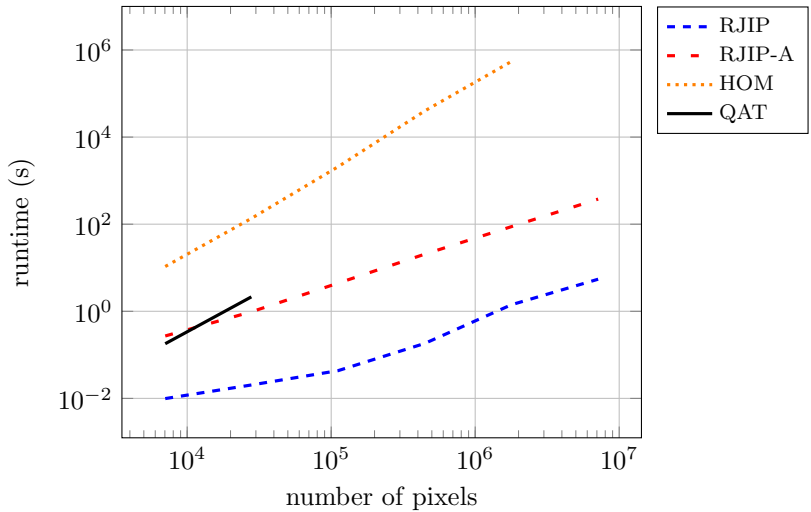
\begin{figure}[t]
\centering

 \begin{tikzpicture}     
\begin{axis}[
xmode=log,
ymode=log,
width=0.7\textwidth,
ylabel={runtime (s)},
xlabel={number of pixels},
xlabel near ticks,
ylabel near ticks,
ymax=10000000,
 y label style={at={(axis description cs:-0.15,.5)},
 anchor=south},
grid = major,
legend entries={
    RJIP ,
    RJIP-A,
    HOM ,   
  QAT},
legend cell align=left,
legend style={legend pos=outer north east, 
font=\footnotesize},
]
\addplot+[mark=none, dashed, very thick, blue]
table[x index=0, y index=7] 
{multiscale_rjip.dat};
\addplot+[mark=none, loosely dashed, very thick, red]
table[x index=0, y index=5] 
{multiscale_rjip_aniso.dat};
\addplot+[mark=none, dotted, very thick, orange]
table[x index=0, y index=9] 
{multiscale_hom.dat};
\addplot+[mark=none, solid, very thick, black]
table[x index=0, y index=9] 
{multiscale_echo.dat};
\end{axis}
\end{tikzpicture} %

 \caption{We compare the runtime scaling behaviour of our pipelines 
 with homogeneous diffusion on five downsampled versions of frame 917 
 of the 4K Cinema\-Scope movie \emph{Sintel} ($4096 \times 1744$).
 Here, both axes are presented in a log scale. The experiments are 
 conducted on a single core of an Intel Core i7-6700A@3.40GHz with 
 32 GB RAM.}
  \label{fig:timing}
\end{figure}

Our experiments consider a single compression with a fixed sampling distance 
$r=4$ ($\approx 4\%$ mask density) and $q=32$ quantisation levels. 
We implemented a version of our RJIP codec which uses homogeneous diffusion
that is solved through a conjugate gradient scheme. We refer to this 
homogeneous diffusion version of the codec HOM. It is to be noted that
our proposed tonal optimisation is not applicable to HOM, which therefore
uses the simple iterative adjustment scheme.  Note that also the more 
sophisticated tonal optimisation options for homogeneous diffusion from the 
previous section are not directly applicable, since they do not take 
quantisation into account and thus would modify the reconstruction quality.
From Fig.~\ref{fig:timing}, we can see that RJIP is faster than 
HOM by up to five orders of magnitude. The 
compression times for RJIP range from $0.009$s for a $128 \times 55$ image 
to $5.45$s for a 4K image. 

\medskip

In contrast to RJIP, it is unpopular to localise HOM by choosing a finite 
stopping time, since one is usually not willing to accept the quality 
deterioration. Thus, the tonal optimisation is more time-consuming for 
non-localised homogeneous diffusion than for the localised RJIP. Peter et 
al.~\cite{PHNH16} proposed 
an alternative \emph{quantisation-aware tonal optimisation} (QAT) which 
increases the speed at the cost of high memory consumption. We could not test 
QAT for images larger than $512 \times 218$, as our test machine ran out of 
memory. RJIP is still $2$ to $3$ orders of magnitude faster than this 
specialised algorithm.

\medskip

We also implemented a uniform mask codec like RJIP, but improved it with 
anisotropic Shepard inpainting, which we call RJIP-A. In addition to the 
parameter optimisation for the grid size and the quantisation parameter, we 
also run an optimisation to find the contrast parameter $\lambda$ and the
variance of the Gaussians $\sigma$ similar to our anisotropic Shepard codec 
on trees. For tonal optimisation, we fall back to the iterative random walk
method, as a clean optimal solution for the tonal value cannot be computed
here as in the case of RJIP. We also alternate the tonal and parameter
optimisation to further increase quality as in Section~\ref{sec:subdivision}.
Finally, as we need all neighbouring pixels to compute derivatives and
consequently the inpainting at that pixel, we cannot perform joint inpainting
and prediction as used in RJIP. Therefore, we simply compress the pixel
values with LPAQ.

\medskip

We can also observe that RJIP-A is about two orders of magnitude slower 
than RJIP. This results from the higher amount of parameters that need to be 
optimised. In addition to the grid parameter $r$ and the quantisation parameter 
$q$, we also need to determine the two additional parameters $\lambda$ and 
$\sigma$ of the anisotropic Shepard inpainting. Moreover, each individual 
inpainting is
slower due to the computations required to calculate the Gaussian kernel
at each point.  Changing a pixel also requires an update
of the neighbourhood derivative information. Thus, tonal optimisation is slower
than for RJIP. In spite of all the additional computations, we are still
faster than HOM by about one to two orders of magnitude. Furthermore, we
can see that our anisotropic codec is slightly slower than QAT for very
low resolutions, but quickly becomes more efficient as resolution increases.

\begin{figure}
\centering
\begin{tabular}{c c}

\emph{trui} & JPEG  \\
\includegraphics[width = 0.25\textwidth]
		{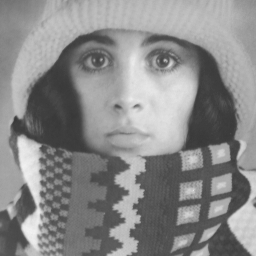} &
		\includegraphics[width = 0.25\textwidth]
		{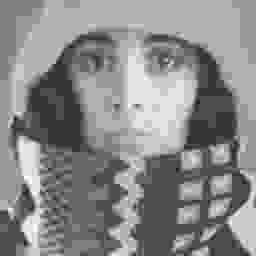} \\
  &  MSE=$160.57$,\, SSIM=$0.738$ \\[2mm]
  HOM & JPEG2000 \\
		\includegraphics[width = 0.25\textwidth]
		{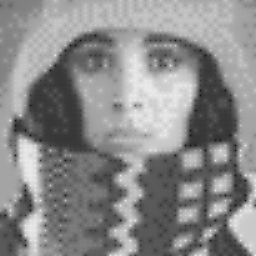} &
		\includegraphics[width = 0.25\textwidth]
		{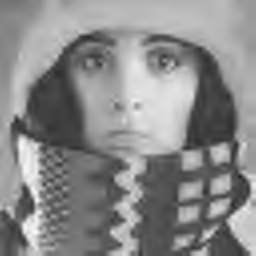} \\
		
		  MSE=$127.38$, \, 
		 SSIM=$0.751$ & MSE=$109.91$,\, SSIM=$0.809$ \\[2mm]
	 
	 RJIP & RJIP-A\\
		\includegraphics[width = 0.25\textwidth]
		{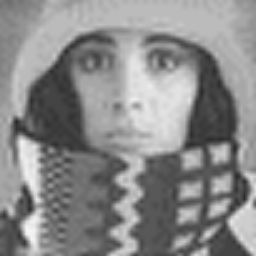} &
		\includegraphics[width = 0.25\textwidth]
		{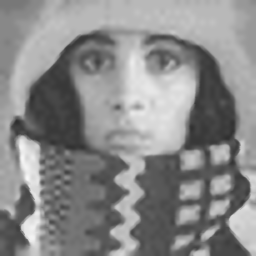} \\
  MSE=$82.86$,\, SSIM=$0.825$ & MSE=$\mathbf{71.50}$,\, 
		SSIM=$\mathbf{0.850}$ \\[2mm]
		\multicolumn{2}{c}{\begin{tikzpicture}     
\begin{axis}[
width=0.45\textwidth, height = 0.34\textwidth,
xmin=0, ymin=0, xmax=160,
ylabel=\small{MSE},
xlabel=\small{compression ratio},
grid = major,
legend entries={
    RJIP,
    HOM,
    JPEG,
    JPEG2000,
    RJIP-A},
legend cell align=left,
legend style=
{legend pos=outer north east, font=\footnotesize},
]
\addplot+[mark=none, dashed, very thick, blue]
table[x index=0, y index=1] {trui_rjip.tab};
\addplot+[mark=none, dotted, very thick, orange]
table[x index=0, y index=1] {trui_hom.dat};
\addplot+[mark=none, solid, very thick, black]
table[x index=0, y index=1] {trui_jpeg.dat};
\addplot+[mark=none, dashdotted, very thick, violet]
table[x index=0, y index=1] {trui_jp2.dat};
\addplot+[mark=none, loosely dashed, very thick, red]
table[x index=0, y index=1] {trui_rjip_aniso_2.tab};

\end{axis}
\end{tikzpicture}} \\
		
		 \multicolumn{2}{c}{Comparisons 
		on the \emph{trui} image} 

\end{tabular}

\caption{We compressed $trui$ with different compression 
methods at a ratio of 70:1 and compare them with the MSE 
error measure. Additionally, we display the SSIM~\cite{WBSS04} 
scores for the presented images. We can see that our Shepard 
inpainting-based methods do not present any unpleasant 
artefacts. From the rate-distortion curves, we can observe 
that our anisotropic Shepard codec with uniform masks 
outperforms JPEG2000 over most compression ratios while 
the base RJIP codec outperforms JPEG2000 at a compression 
ratio of 60.}
\label{fig:trui}
\end{figure}

\subsection{Comparing Uniform Mask Codecs}

In this set of comparisons, we compare the base RJIP codec with isotropic 
Shepard inpainting on uniform masks to RJIP-A, the uniform
mask codec with anisotropic Shepard inpainting. This allows us to evaluate 
the relative performance of the anisotropic Shepard operator on the piecewise 
smooth test image \emph{trui} image and the 500 textured images of the Berkeley 
dataset~\cite{MFTM01}. Moreover, we also compare against JPEG and JPEG2000.

\medskip

From Fig.~\ref{fig:trui}, we observe that on the piecewise smooth \emph{trui} 
image, RJIP-A performs the best across
almost all compression ratios, even compared to JPEG2000. It benefits from its 
superior reconstruction of directional structures with anisotropic Shepard 
inpainting.

\medskip

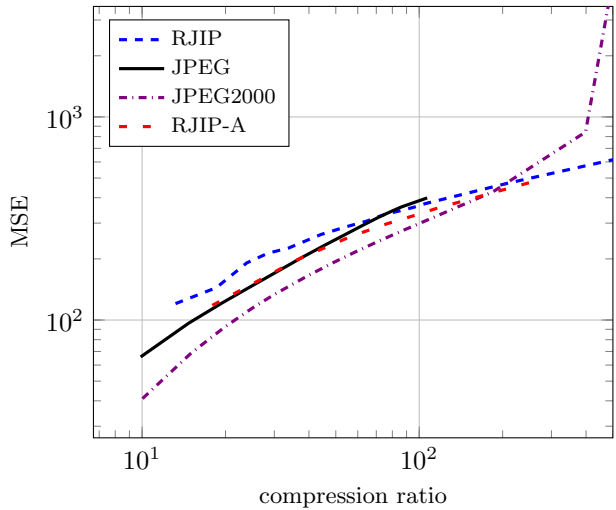
\begin{figure}[t]
 \centering
    \begin{tikzpicture}     
    \begin{axis}[
    xmode=log,
    ymode=log,
    log basis x =10,
    width=0.7\textwidth,
    ymax=3500,
    xmax=500,
    restrict x to domain=0:1000,
    ylabel=\small{MSE},
    xlabel=\small{compression ratio},
    xlabel near ticks,
    ylabel near ticks,
    grid = major,
    legend entries={
        RJIP,
        JPEG,           
        JPEG2000,
        RJIP-A},
    legend cell align=left,
    legend style={legend pos=north west, font=\footnotesize},
    ]
    \addplot+[mark=none, dashed, very thick, blue]
    table[x index=0, y index=1] {berkeley_rjip.tab};

    \addplot+[mark=none, solid, very thick, black]
    table[x index=0, y index=1] {berkeley_jpg.tab};
    \addplot+[mark=none, dashdotted, very thick, violet]
    table[x index=0, y index=1] {berkeley_jp2.tab};
    \addplot+[mark=none, loosely dashed, very thick, red]
    table[x index=0, y index=1] {berkeley_grey_rjip_aniso.csv};
    \end{axis}
    \end{tikzpicture}

\caption{ On the Berkeley database with textured images, RJIP-A 
is competitive to JPEG at low compression ratios whereas RJIP and 
RJIP-A outperforms JPEG and JPEG2000 for high compression ratios. 
Both axes in the plot are presented in a log scale.}
\label{fig:berkeley}
\end{figure}

For the Berkeley dataset, this advantage is also visible for low to medium 
compression ratios in Fig.~\ref{fig:berkeley}: RJIP-A outperforms its isotropic 
counterpart RJIP and is competitive with JPEG. At higher compression ratios, both 
Shepard codecs yield very similar quality and are able to beat both JPEG and 
JPEG2000. These very sparse mask grids contain less reliable information on 
anisotropy, which can cause the anisotropic Gaussian kernels to degenerate to 
an isotropic setting. 

\medskip

The more favourable rate-distortion behaviour at higher compression ratios 
compared to transform-based compression results from the fact that the number 
of mask points to be stored does not need to be reduced proportionally to the 
compression ratio. At high ratios, coarse quantisation can be used to reduce 
the coding cost, and the smooth inpainting is able to restore a wider range of 
grey values than that of the mask pixels. This makes our codecs particularly 
suited for compressing images at high compression ratios. 	

\subsection{Comparing Subdivision-Based Codecs}

In our final comparison, we consider the rate-distortion behaviour of our 
subdivision codecs for isotropic and anisotropic Shepard inpainting on 
the greyscale Kodak dataset. We also include the uniform mask codecs 
from the previous set of experiments and consider JPEG as a reference.

\medskip

Our uniform mask codecs use multiple target compression ratios
while we specify different target splitting errors for our subdivision codecs 
to steer the rate-distortion trade-off. For this set of experiments, we present 
images that show representative performance for different image content and 
different behaviour of our algorithms in Fig.~\ref{fig:final_results}. The rate-
distortion curves for all images in the Kodak dataset can be found in
the supplementary material.

\medskip

Fig.~\ref{fig:final_results} illustrates that for images with a
dominant foreground and a homogeneous background, the subdivision codecs
outperform their regular grid counterparts. In such images,
the subdivision codecs benefit from denser known data in textured regions. This 
also leads to a higher accuracy of the derivative
approximations for anisotropic Shepard inpainting in such dense regions,
yielding a better overall reconstruction quality. Even at higher
compression ratios, our subdivision codecs can retain more structures compared
to the original RJIP codec. 

\medskip

On the other hand, for images that are highly textured overall, the advantages 
of the subdivision codec diminish. Since all regions are similarly detailed, it 
produces uniformly distributed masks and thus degenerates to the uniform codec. 
However, it still requires significant overhead for the tree structure and thus 
performs worse than its simpler counterpart. This suggests switching between 
both coding archetypes depending on the image content.

 \begin{figure}
 	\centering
 	\begin{tabular}{c c}
 
 	\includegraphics[width = 0.33\textwidth]
 	{kodim23} & 
 	\includegraphics[width = 0.33\textwidth]
 	{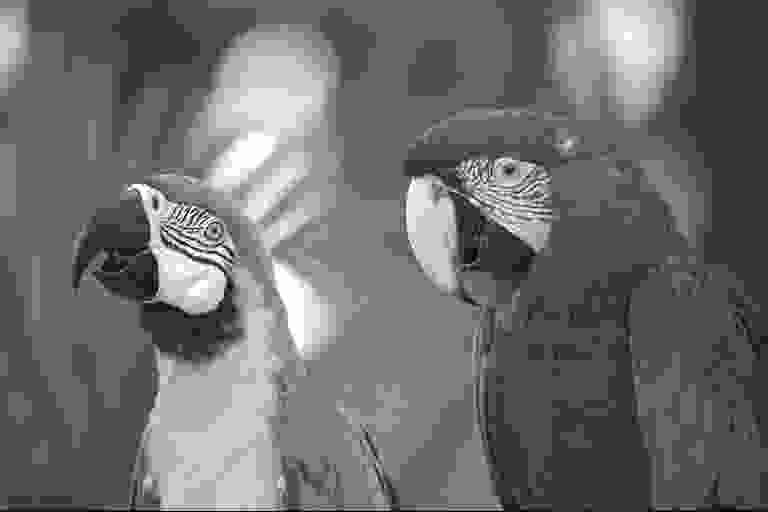} \\
  \emph{kodim23} & JPEG \\
  & ratio 116:1, MSE = 124.63 \\
 	\includegraphics[width = 0.33\textwidth]
 	{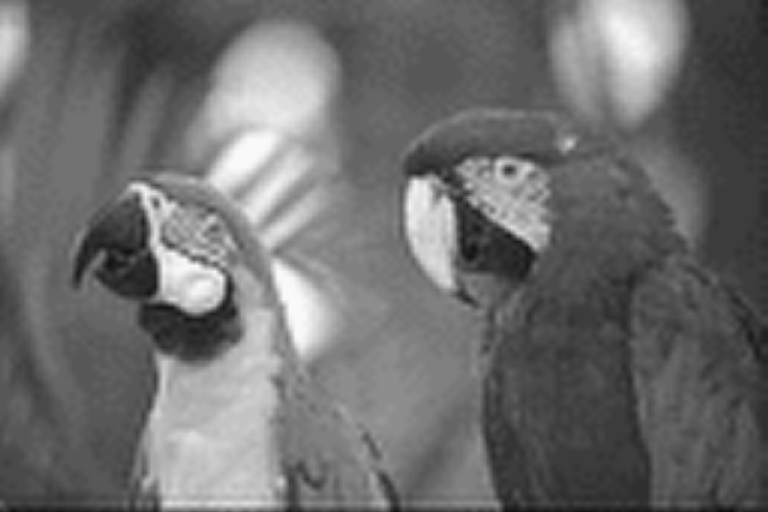} & 
 	\includegraphics[width = 0.33\textwidth]
 	{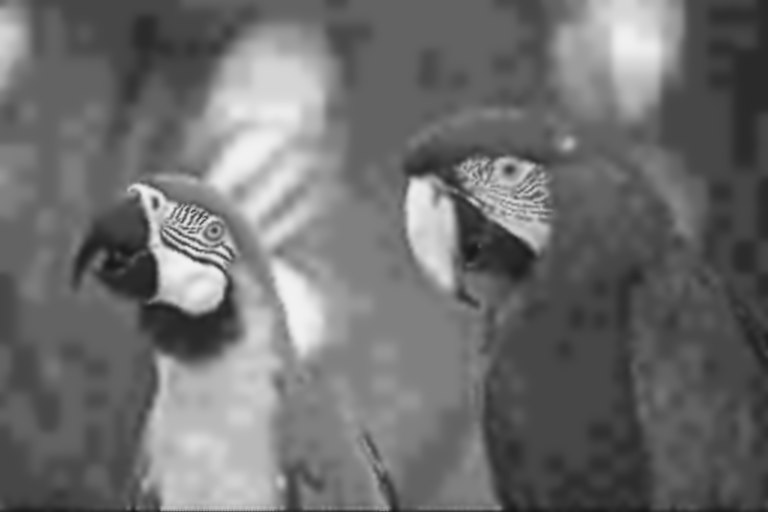} 
 	\\
 		
 	  uniform isotropic (RJIP) & 
 	tree isotropic  \\
 	ratio 117:1, MSE = 113.06 & 
 	ratio 116:1, MSE = 97.82\\
 	\includegraphics[width = 0.33\textwidth]
 	{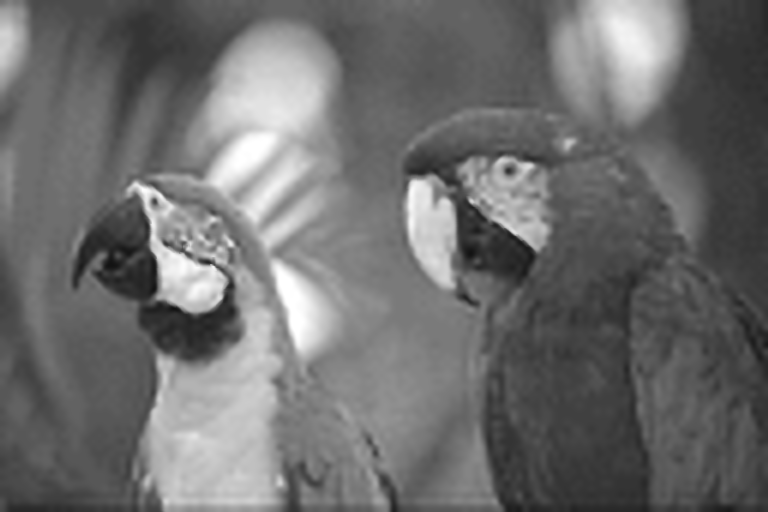} &
 	\includegraphics[width = 0.33\textwidth]
 	{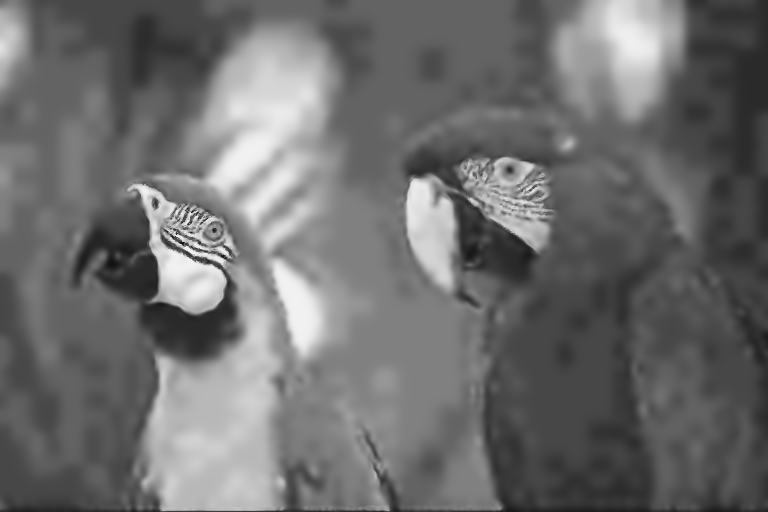} \\
  uniform anisotropic (RJIP-A) & tree anisotropic \\
  ratio {109:1}, MSE = {107.10}  & ratio 120:1, 
 MSE = \textbf{83.93} \\
  
     \multicolumn{2}{c}
 	{
 	\begin{tikzpicture}
 	\begin{axis}[height=0.34\textwidth, width=0.45\textwidth,
 	xlabel={ {compression ratio}},
 	ylabel={ {MSE}},
 	grid = major,
 			legend entries={{}{RJIP},
 		{}{RJIP-A},
 		{}{isotropic$,$ tree},
 	{}{anisotropic$,$ tree},
 	JPEG},
 	legend cell align=left,
  legend style={legend pos=outer north east, font=\footnotesize}
 ]
 \addplot+[mark=none, dashed, very thick, blue] table[x index=6, 
 	y index=7, 
 	select coords between index={132}{137}, col sep = tab] 
 	{final_data.csv};
 	\addplot+[mark=none, loosely dashed, very thick, red] 
 	table[x index=12, y index=13, 
 	select coords between index={132}{137}, col sep = tab] 
 	{final_data.csv};
 	\addplot+[mark=none, dotted, very thick, orange] 
 	table[x index=3, y index=4, 
 	select coords between index={132}{137}, col sep = tab] 
 	{final_data.csv};
 		\addplot+[mark=none, dashdotted, very thick, purple] 
 		table[x index=0, y index=1, 
 	select coords between index={132}{137}, col sep = tab] 
 	{final_data.csv};
 		\addplot+[mark=none, solid, very thick, black] 
 		table[x index=9, y index=10, 
 	select coords between index={132}{137}, col sep = tab] 
 	{final_data.csv};
 	\end{axis}
 	\end{tikzpicture}
 	}
 	
  \\
   
  \multicolumn{2}{c}{rate-distortion 
  comparisons for \emph{kodim23}} \\

 		\end{tabular}
 		
 \end{figure}

 \begin{figure}\ContinuedFloat
 	\centering
 	\begin{tabular}{c c}
 
 \includegraphics[width = 0.33\textwidth]
 {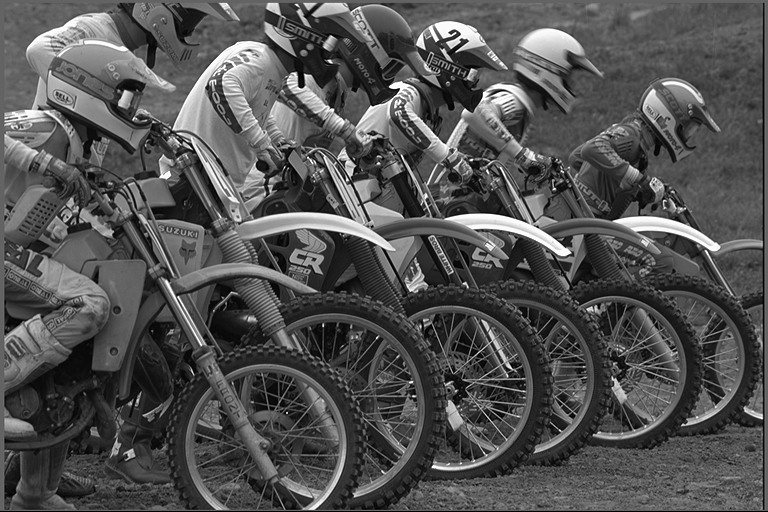} & 
 \includegraphics[width = 0.33\textwidth]
 {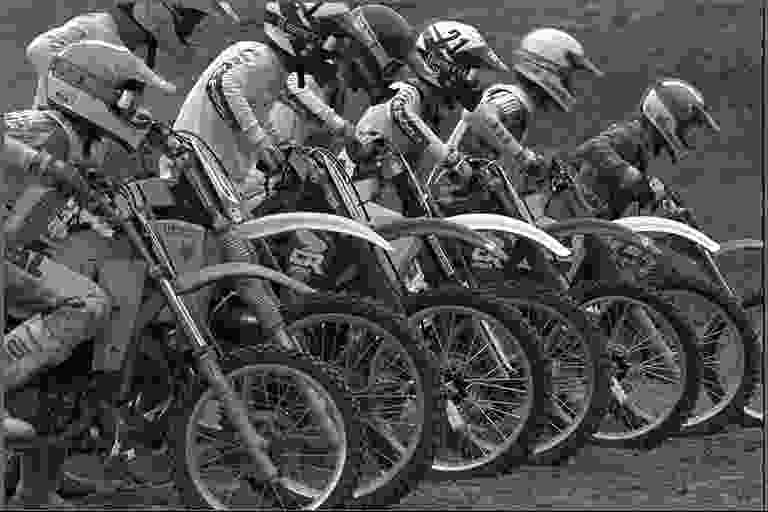} \\
 \emph{kodim05} & JPEG \\
 &  ratio 37:1, MSE = 357.34 \\
 \includegraphics[width = 0.33\textwidth]
 {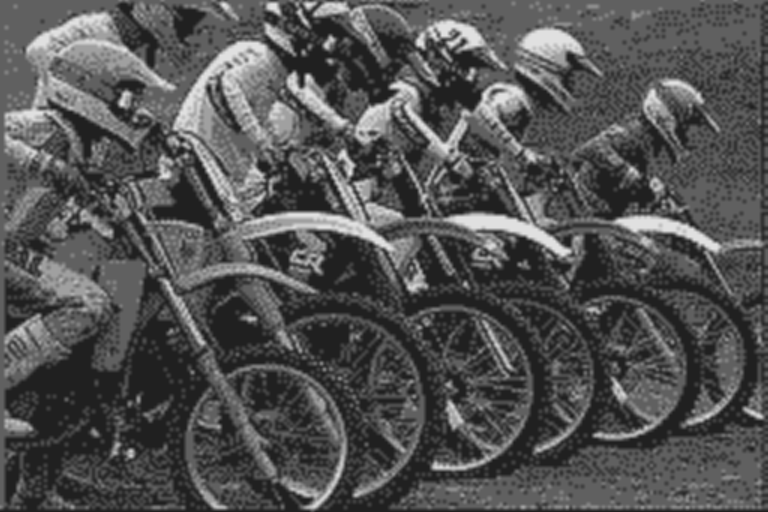} & 
 \includegraphics[width = 0.33\textwidth]
 {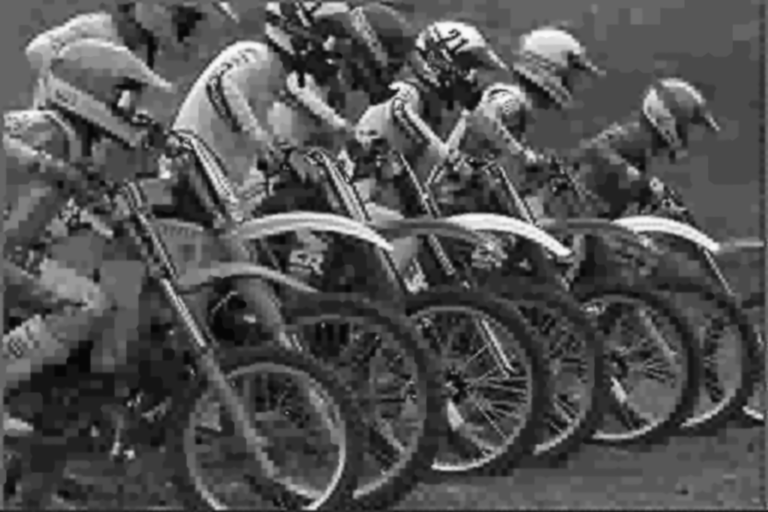} 
 \\
 		
  uniform isotropic (RJIP)  & tree isotropic\\
  ratio 46:1, MSE = 423.26  
 & ratio 41:1, MSE = 433.64\\

 \includegraphics[width = 0.33\textwidth]
 {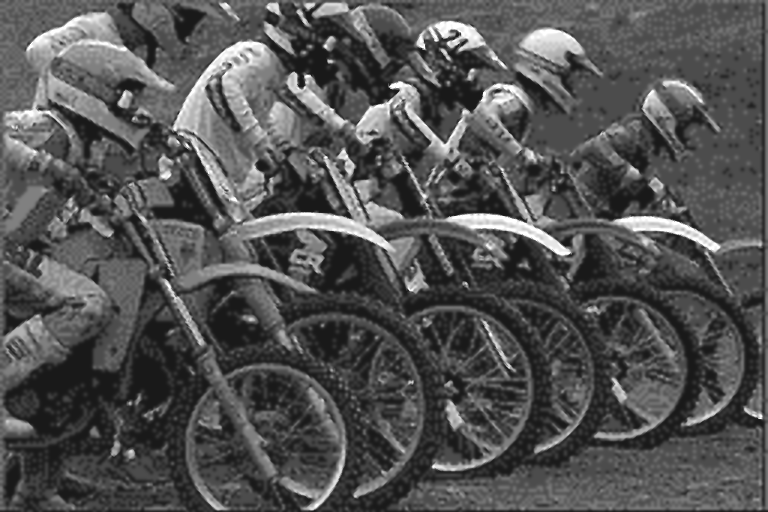} & 
 \includegraphics[width = 0.33\textwidth]
 {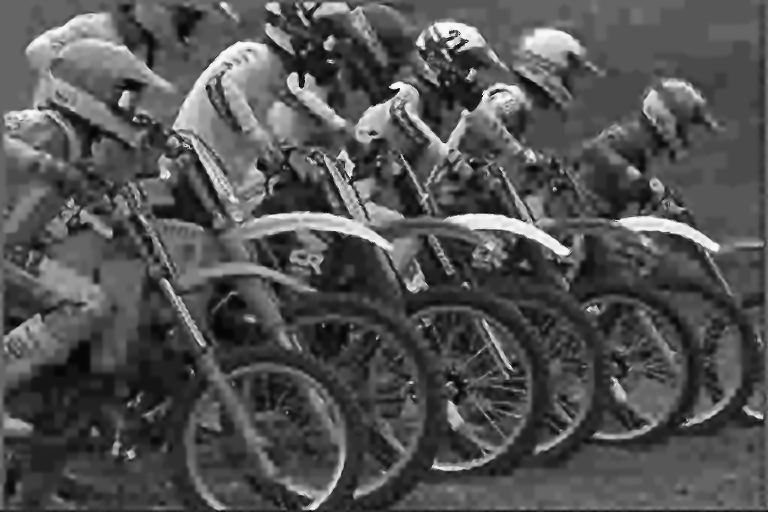} \\
 uniform anisotropic 
 (RJIP-A) & tree anisotropic \\
 ratio 41.56:1, MSE = \textbf{316.24} & ratio 42:1, MSE = 401.38 \\
 \multicolumn{2}{c}{
 	\begin{tikzpicture}
 	\begin{axis}[height=0.34\textwidth, width=0.45\textwidth,
 	xlabel={ {compression ratio}},
 	ylabel={ {MSE}},
 	grid = major,
 		legend entries={{}{RJIP},
 		{}{RJIP-A},
 		{}{isotropic$,$ tree},
 	{}{anisotropic$,$ tree},
 	JPEG},
 	legend cell align=left,
  legend style={legend pos=outer north east, font = \footnotesize}
 ]
 	\addplot+[mark=none, dashed, very thick, blue] 
 	table[x index=6, y index=7, 
 	select coords between index={24}{29}, col sep = tab] 
 	{final_data.csv};
 	\addplot+[mark=none, loosely dashed, very thick, red] 
 	table[x index=12, y index=13, 
 	select coords between index={24}{29}, col sep = tab] 
 	{final_data.csv};
 	\addplot+[mark=none, dotted, very thick, orange] 
 	table[x index=3, y index=4, 
 	select coords between index={24}{29}, col sep = tab] 
 	{final_data.csv};
 		\addplot+[mark=none, dashdotted, very thick, purple] 
 		table[x index=0, y index=1, 
 	select coords between index={24}{29}, col sep = tab] 
 	{final_data.csv};
 		\addplot+[mark=none, solid, very thick, black] 
 		table[x index=9, y index=10, 
 	select coords between index={24}{29}, col sep = tab] 
 	{final_data.csv};
 	\end{axis}
 	\end{tikzpicture}
 	}
  \\
 	    
\multicolumn{2}{c}
 {rate-distortion comparisons for \emph{kodim05}} \\
 
 		\end{tabular}
 		
 \caption{Our anisotropic Shepard inpainting-based codecs 
 perform better than their isotropic counterparts overall 
 and outperform JPEG at higher compression ratios. From the 
 plots, we can see that the subdivision-based codecs perform 
 better than the regular mask-based codecs if the image has 
 a clear foreground subject. This is caused by the fact that they can 
 adapt the mask such that the detailed regions are reconstructed 
 better. On the other hand, the regular mask-based codecs perform 
 better if the image is textured overall.}
 \label{fig:final_results}
 \end{figure}
 	
\section{Conclusions and Outlook}
\label{sec:conclusion}

We have presented a family of simple but efficient image compression methods 
that are based on Shepard inpainting. It illustrates that straightforward
inpainting methods can become competitive if they are equipped with carefully 
chosen components. Joint inpainting and prediction is a novel concept that increases coding 
efficiency at no additional storage overhead. While it thrives in combination 
with very efficient inpainting approaches such as Shepard interpolation, it can 
also be integrated in other future inpainting-based approaches.

\medskip

Furthermore, our anisotropic Shepard extension demonstrates that sharp 
reconstructions from sparse data are possible at fast runtimes that are so far 
only associated with sophisticated implementations of linear diffusion 
inpainting. Due to its ease of implementation, anisotropic Shepard inpainting 
offers an alternative for future practical applications. In particular, for 
images with clearly separated fore- and background, it performs well in 
combination with our adaptive subdivision schemes.

\medskip

In general, the straightforward Shepard codecs are competitive with existing 
transform-based codecs for high compression ratios. They do not produce block 
artefacts even with very little stored data and can also outperform both JPEG 
and JPEG2000 quantitatively, especially on piecewise-smooth images.

\medskip

In the future, we aim to extend our Shepard-based codecs to colour images by 
employing dedicated strategies that exploit human
perception. As Shepard inpainting excels on piecewise smooth data, we hope
that it contributes to the compression of depth maps or flow fields. First
attempts that followed our earlier conference publication~\cite{Pe19} already
yielded promising results~\cite{JPW21}. 

\subsection*{Acknowledgements}

We thank Vassillen Chizhov for providing his implementation of Voronoi
decomposition, which was integrated into the subdivision-based anisotropic
Shepard codec. We also thank Dr Matthias Augustin for his implementation
of the Green's functions implementation of tonal optimisation for homogeneous
diffusion~\cite{Ho16}, which was used for comparisons in 
Section~\ref{sec:inp_comp}.  
 \FloatBarrier

\bibliographystyle{elsarticle-num}
\bibliography{bib}

\clearpage

\begin{figure*}[ht]
	\centering
	\begin{tabular}{c c c}
	\multicolumn{3}{c}{\large{\textbf{ Supplementary Material}}}   \\[7mm]
	 \multicolumn{3}{c}{
 \begin{tikzpicture}
\begin{customlegend}
[legend columns=3,legend style={align=center,draw=none},
        legend entries={{RJIP} ,
                        {RJIP-A} ,
                        {isotropic, subdivision}
                        }]
        \addlegendimage{mark=none, dashed, very thick, blue}
        \addlegendimage{mark=none, loosely dashed, very thick, red}
        \addlegendimage{mark=none, dotted, very thick, orange}
        
        \end{customlegend}
\end{tikzpicture}
	    } \\

     	 \multicolumn{3}{c}{
 \begin{tikzpicture}
\begin{customlegend}
[legend columns=2,legend style={align=center,draw=none},
        legend entries={
                        {anisotropic, subdivision},
                        {JPEG}
                        }]
        
        \addlegendimage{mark=none, dashdotted, very thick, purple}
        \addlegendimage{mark=none, solid, very thick, black}
        \end{customlegend}
\end{tikzpicture}
	    }
	    
	    \\[3mm]
	\begin{tikzpicture}
	\begin{axis}[height=0.3\textwidth, width=0.32\textwidth,
	xlabel={ {compression ratio}},
	ylabel={ {MSE}},
	grid = major
]
\addplot+[mark=none, dashed, very thick, blue] 
table[x index=6, y index=7, 
select coords between index={0}{5}, col sep = tab]
{final_data.csv};
\addplot+[mark=none, loosely dashed, very thick, red] 
table[x index=12, y index=13, 
select coords between index={0}{5}, col sep = tab] 
{final_data.csv};
\addplot+[mark=none, dotted, very thick, orange] 
table[x index=3, y index=4, 
select coords between index={0}{5}, col sep = tab] 
{final_data.csv};
\addplot+[mark=none, dashdotted, very thick, purple] 
table[x index=0, y index=1, 
select coords between index={0}{5}, col sep = tab] 
{final_data.csv};
\addplot+[mark=none, solid, very thick, black] 
table[x index=9, y index=10, 
select coords between index={0}{5}, col sep = tab] 
{final_data.csv};
\end{axis}
\end{tikzpicture}
&
\begin{tikzpicture}
\begin{axis}[height=0.3\textwidth, width=0.32\textwidth,
xlabel={ {compression ratio}},
ylabel={ {MSE}},
grid = major
]
	\addplot+[mark=none, dashed, very thick, blue] 
	table[x index=6, y index=7, 
	select coords between index={6}{11}, col sep = tab] 
	{final_data.csv};
	\addplot+[mark=none, loosely dashed, very thick, red] 
	table[x index=12, y index=13, 
	select coords between index={6}{11}, col sep = tab] 
	{final_data.csv};
	\addplot+[mark=none, dotted, very thick, orange] 
	table[x index=3, y index=4, 
	select coords between index={6}{11}, col sep = tab]
	{final_data.csv};
		\addplot+[mark=none, dashdotted, very thick, purple] 
		table[x index=0, y index=1, 
	select coords between index={6}{11}, col sep = tab] 
	{final_data.csv};
		\addplot+[mark=none, solid, very thick, black]
		table[x index=9, y index=10, 
	select coords between index={6}{11}, col sep = tab]
	{final_data.csv};
	\end{axis}
	\end{tikzpicture}
	&
		\begin{tikzpicture}
	\begin{axis}[height=0.3\textwidth, width=0.32\textwidth,
	xlabel={ {compression ratio}},
	ylabel={ {MSE}},
	grid = major
]
	\addplot+[mark=none, dashed, very thick, blue] 
	table[x index=6, y index=7, 
	select coords between index={12}{17}, col sep = tab] 
	{final_data.csv};
	\addplot+[mark=none, loosely dashed, very thick, red] 
	table[x index=12, y index=13, 
	select coords between index={12}{17}, col sep = tab] 
	{final_data.csv};
	\addplot+[mark=none, dotted, very thick, orange] 
	table[x index=3, y index=4, 
	select coords between index={12}{17}, col sep = tab] 
	{final_data.csv};
		\addplot+[mark=none, dashdotted, very thick, purple]
		table[x index=0, y index=1, 
	select coords between index={12}{17}, col sep = tab] 
	{final_data.csv};
		\addplot+[mark=none, solid, very thick, black] 
		table[x index=9, y index=10, 
	select coords between index={12}{17}, col sep = tab] 
	{final_data.csv};
	\end{axis}
	\end{tikzpicture}
	\\
	\emph{kodim01} & \emph{kodim02} & \emph{kodim03} \\[0.4cm]
	
	\begin{tikzpicture}
	\begin{axis}[height=0.3\textwidth, width=0.32\textwidth,
	xlabel={ {compression ratio}},
	ylabel={ {MSE}},
	grid = major
]
	\addplot+[mark=none, dashed, very thick, blue] 
	table[x index=6, y index=7, 
	select coords between index={18}{23}, col sep = tab] 
	{final_data.csv};
	\addplot+[mark=none, loosely dashed, very thick, red] 
	table[x index=12, y index=13, 
	select coords between index={18}{23}, col sep = tab] 
	{final_data.csv};
	\addplot+[mark=none, dotted, very thick, orange] 
	table[x index=3, y index=4, 
	select coords between index={18}{23}, col sep = tab] 
	{final_data.csv};
		\addplot+[mark=none, dashdotted, very thick, purple] 
		table[x index=0, y index=1, 
	select coords between index={18}{23}, col sep = tab] 
	{final_data.csv};
		\addplot+[mark=none, solid, very thick, black] 
		table[x index=9, y index=10, 
	select coords between index={18}{23}, col sep = tab] 
	{final_data.csv};
	\end{axis}
	\end{tikzpicture} &
	
	\begin{tikzpicture}
	\begin{axis}[height=0.3\textwidth, width=0.32\textwidth,
	xlabel={ {compression ratio}},
	ylabel={ {MSE}},
	grid = major
]
	\addplot+[mark=none, dashed, very thick, blue] 
	table[x index=6, y index=7, 
	select coords between index={24}{29}, col sep = tab] 
	{final_data.csv};
	\addplot+[mark=none, loosely dashed, very thick, red] 
	table[x index=12, y index=13, 
	select coords between index={24}{29}, col sep = tab] 
	{final_data.csv};
	\addplot+[mark=none, dotted, very thick, orange] 
	table[x index=3, y index=4, 
	select coords between index={24}{29}, col sep = tab] 
	{final_data.csv};
		\addplot+[mark=none, dashdotted, very thick, purple] 
		table[x index=0, y index=1, 
	select coords between index={24}{29}, col sep = tab] 
	{final_data.csv};
		\addplot+[mark=none, solid, very thick, black] 
		table[x index=9, y index=10, 
	select coords between index={24}{29}, col sep = tab] 
	{final_data.csv};
	\end{axis}
	\end{tikzpicture} & 
	\begin{tikzpicture}
	\begin{axis}[height=0.3\textwidth, width=0.32\textwidth,
	xlabel={ {compression ratio}},
	ylabel={ {MSE}},
	grid = major
]
	\addplot+[mark=none, dashed, very thick, blue] 
	table[x index=6, y index=7, 
	select coords between index={30}{35}, col sep = tab] 
	{final_data.csv};
	\addplot+[mark=none, loosely dashed, very thick, red] 
	table[x index=12, y index=13, 
	select coords between index={30}{35}, col sep = tab] 
	{final_data.csv};
	\addplot+[mark=none, dotted, very thick, orange] 
	table[x index=3, y index=4, 
	select coords between index={30}{35}, col sep = tab] 
	{final_data.csv};
		\addplot+[mark=none, dashdotted, very thick, purple] 
		table[x index=0, y index=1, 
	select coords between index={30}{35}, col sep = tab] 
	{final_data.csv};
		\addplot+[mark=none, solid, very thick, black] 
		table[x index=9, y index=10, 
	select coords between index={30}{35}, col sep = tab] 
	{final_data.csv};
	\end{axis}
	\end{tikzpicture}\\
	
	\emph{kodim04} & \emph{kodim05} & \emph{kodim06} \\[0.4cm]
	
			\begin{tikzpicture}
	\begin{axis}[height=0.3\textwidth, width=0.32\textwidth,
	xlabel={ {compression ratio}},
	ylabel={ {MSE}},
	grid = major
]
	\addplot+[mark=none, dashed, very thick, blue] 
	table[x index=6, y index=7, 
	select coords between index={36}{41}, col sep = tab] 
	{final_data.csv};
	\addplot+[mark=none, loosely dashed, very thick, red]
	table[x index=12, y index=13, 
	select coords between index={36}{41}, col sep = tab]
	{final_data.csv};
	\addplot+[mark=none, dotted, very thick, orange]
	table[x index=3, y index=4, 
	select coords between index={36}{41}, col sep = tab] 
	{final_data.csv};
		\addplot+[mark=none, dashdotted, very thick, purple]
		table[x index=0, y index=1, 
	select coords between index={36}{41}, col sep = tab]
	{final_data.csv};
		\addplot+[mark=none, solid, very thick, black]
		table[x index=9, y index=10, 
	select coords between index={36}{41}, col sep = tab]
	{final_data.csv};
	\end{axis}
	\end{tikzpicture} &
	
	\begin{tikzpicture}
	\begin{axis}[height=0.3\textwidth, width=0.32\textwidth,
	xlabel={ {compression ratio}},
	ylabel={ {MSE}},
	grid = major
]
	\addplot+[mark=none, dashed, very thick, blue]
	table[x index=6, y index=7, 
	select coords between index={42}{47}, col sep = tab]
	{final_data.csv};
	\addplot+[mark=none, loosely dashed, very thick, red]
	table[x index=12, y index=13, 
	select coords between index={42}{47}, col sep = tab]
	{final_data.csv};
	\addplot+[mark=none, dotted, very thick, orange]
	table[x index=3, y index=4, 
	select coords between index={42}{47}, col sep = tab]
	{final_data.csv};
		\addplot+[mark=none, dashdotted, very thick, purple]
		table[x index=0, y index=1, 
	select coords between index={42}{47}, col sep = tab]
	{final_data.csv};
		\addplot+[mark=none, solid, very thick, black]
		table[x index=9, y index=10, 
	select coords between index={42}{47}, col sep = tab]
	{final_data.csv};
	\end{axis}
	\end{tikzpicture} & 
	\begin{tikzpicture}
	\begin{axis}[height=0.3\textwidth, width=0.32\textwidth,
	xlabel={ {compression ratio}},
	ylabel={ {MSE}},
	grid = major
]
	\addplot+[mark=none, dashed, very thick, blue]
	table[x index=6, y index=7, 
	select coords between index={48}{53}, col sep = tab]
	{final_data.csv};
	\addplot+[mark=none, loosely dashed, very thick, red]
	table[x index=12, y index=13, 
	select coords between index={48}{53}, col sep = tab]
	{final_data.csv};
	\addplot+[mark=none, dotted, very thick, orange] 
	table[x index=3, y index=4, 
	select coords between index={48}{53}, col sep = tab] 
	{final_data.csv};
		\addplot+[mark=none, dashdotted, very thick, purple] 
		table[x index=0, y index=1, 
	select coords between index={48}{53}, col sep = tab] 
	{final_data.csv};
		\addplot+[mark=none, solid, very thick, black]
		table[x index=9, y index=10, 
	select coords between index={48}{53}, col sep = tab] 
	{final_data.csv};
	\end{axis}
	\end{tikzpicture}\\
	
	\emph{kodim07} & \emph{kodim08} & \emph{kodim09} \\[0.4cm]
	
	\begin{tikzpicture}
	\begin{axis}[height=0.3\textwidth, width=0.32\textwidth,
	xlabel={ {compression ratio}},
	ylabel={ {MSE}},
	grid = major
]
	\addplot+[mark=none, dashed, very thick, blue] 
	table[x index=6, y index=7, 
	select coords between index={54}{59}, col sep = tab]
	{final_data.csv};
	\addplot+[mark=none, loosely dashed, very thick, red] 
	table[x index=12, y index=13, 
	select coords between index={54}{59}, col sep = tab] 
	{final_data.csv};
	\addplot+[mark=none, dotted, very thick, orange] 
	table[x index=3, y index=4, 
	select coords between index={54}{59}, col sep = tab] 
	{final_data.csv};
		\addplot+[mark=none, dashdotted, very thick, purple] 
		table[x index=0, y index=1, 
	select coords between index={54}{59}, col sep = tab] 
	{final_data.csv};
		\addplot+[mark=none, solid, very thick, black]
		table[x index=9, y index=10, 
	select coords between index={54}{59}, col sep = tab]
	{final_data.csv};
	\end{axis}
	\end{tikzpicture} &
	
	\begin{tikzpicture}
	\begin{axis}[height=0.3\textwidth, width=0.32\textwidth,
	xlabel={ {compression ratio}},
	ylabel={ {MSE}},
	grid = major
]
	\addplot+[mark=none, dashed, very thick, blue]
	table[x index=6, y index=7, 
	select coords between index={60}{65}, col sep = tab]
	{final_data.csv};
	\addplot+[mark=none, loosely dashed, very thick, red]
	table[x index=12, y index=13, 
	select coords between index={60}{65}, col sep = tab] 
	{final_data.csv};
	\addplot+[mark=none, dotted, very thick, orange] 
	table[x index=3, y index=4, 
	select coords between index={60}{65}, col sep = tab] 
	{final_data.csv};
		\addplot+[mark=none, dashdotted, very thick, purple]
		table[x index=0, y index=1, 
	select coords between index={60}{65}, col sep = tab]
	{final_data.csv};
		\addplot+[mark=none, solid, very thick, black]
		table[x index=9, y index=10, 
	select coords between index={60}{65}, col sep = tab]
	{final_data.csv};
	\end{axis}
	\end{tikzpicture} & 
	\begin{tikzpicture}
	\begin{axis}[height=0.3\textwidth, width=0.32\textwidth,
	xlabel={ {compression ratio}},
	ylabel={ {MSE}},
	grid = major
]
	\addplot+[mark=none, dashed, very thick, blue]
	table[x index=6, y index=7, 
	select coords between index={66}{71}, col sep = tab]
	{final_data.csv};
	\addplot+[mark=none, loosely dashed, very thick, red] 
	table[x index=12, y index=13, 
	select coords between index={66}{71}, col sep = tab]
	{final_data.csv};
	\addplot+[mark=none, dotted, very thick, orange] 
	table[x index=3, y index=4, 
	select coords between index={66}{71}, col sep = tab]
	{final_data.csv};
	\addplot+[mark=none, dashdotted, very thick, purple]
		table[x index=0, y index=1, 
	select coords between index={66}{71}, col sep = tab]
	{final_data.csv};
	\addplot+[mark=none, solid, very thick, black] 
		table[x index=9, y index=10, 
	select coords between index={66}{71}, col sep = tab] 
	{final_data.csv};
	\end{axis}
	\end{tikzpicture}\\
	\emph{kodim10} & \emph{kodim11} & \emph{kodim12} \\[3mm]
	
		\end{tabular}
		
	\caption{Rate-distortion comparisons on the Kodak database}
	\end{figure*}
	
	\begin{figure*}[ht]
	\centering
	\begin{tabular}{c c c}
	 \multicolumn{3}{c}{
 \begin{tikzpicture}
\begin{customlegend}
[legend columns=3,legend style={align=center,draw=none},
        legend entries={{RJIP} ,
                        {RJIP-A} ,
                        {isotropic, subdivision}
                        }]
        \addlegendimage{mark=none, dashed, very thick, blue}
        \addlegendimage{mark=none, loosely dashed, very thick, red}
        \addlegendimage{mark=none, dotted, very thick, orange}
        
        \end{customlegend}
\end{tikzpicture}
	    } \\

     	 \multicolumn{3}{c}{
 \begin{tikzpicture}
\begin{customlegend}
[legend columns=2,legend style={align=center,draw=none},
        legend entries={
                        {anisotropic, subdivision},
                        {JPEG}
                        }]
        
        \addlegendimage{mark=none, dashdotted, very thick, purple}
        \addlegendimage{mark=none, solid, very thick, black}
        \end{customlegend}
\end{tikzpicture}
	    }
	    
	    \\[3mm]
	    
	\begin{tikzpicture}
	\begin{axis}[height=0.3\textwidth, width=0.32\textwidth,
	xlabel={ {compression ratio}},
	ylabel={ {MSE}},
	grid = major
]
	\addplot+[mark=none, dashed, very thick, blue]
	table[x index=6, y index=7, 
	select coords between index={72}{77}, col sep = tab]
	{final_data.csv};
	\addplot+[mark=none, loosely dashed, very thick, red]
	table[x index=12, y index=13, 
	select coords between index={72}{77}, col sep = tab]
	{final_data.csv};
	\addplot+[mark=none, dotted, very thick, orange]
	table[x index=3, y index=4, 
	select coords between index={72}{77}, col sep = tab] 
	{final_data.csv};
		\addplot+[mark=none, dashdotted, very thick, purple] 
		table[x index=0, y index=1, 
	select coords between index={72}{77}, col sep = tab]
	{final_data.csv};
		\addplot+[mark=none, solid, very thick, black]
		table[x index=9, y index=10, 
	select coords between index={72}{77}, col sep = tab] 
	{final_data.csv};
	\end{axis}
	\end{tikzpicture}
	&
		\begin{tikzpicture}
	\begin{axis}[height=0.3\textwidth, width=0.32\textwidth,
	xlabel={ {compression ratio}},
	ylabel={ {MSE}},
	grid = major
]
	\addplot+[mark=none, dashed, very thick, blue] 
	table[x index=6, y index=7, 
	select coords between index={78}{83}, col sep = tab]
	{final_data.csv};
	\addplot+[mark=none, loosely dashed, very thick, red] 
	table[x index=12, y index=13, 
	select coords between index={78}{83}, col sep = tab]
	{final_data.csv};
	\addplot+[mark=none, dotted, very thick, orange] 
	table[x index=3, y index=4, 
	select coords between index={78}{83}, col sep = tab]
	{final_data.csv};
		\addplot+[mark=none, dashdotted, very thick, purple] 
		table[x index=0, y index=1, 
	select coords between index={78}{83}, col sep = tab] 
	{final_data.csv};
		\addplot+[mark=none, solid, very thick, black]
		table[x index=9, y index=10, 
	select coords between index={78}{83}, col sep = tab]
	{final_data.csv};
	\end{axis}
	\end{tikzpicture}
	&
		\begin{tikzpicture}
	\begin{axis}[height=0.3\textwidth, width=0.32\textwidth,
	xlabel={ {compression ratio}},
	ylabel={ {MSE}},
	grid = major
]
	\addplot+[mark=none, dashed, very thick, blue]
	table[x index=6, y index=7, 
	select coords between index={84}{89}, col sep = tab]
	{final_data.csv};
	\addplot+[mark=none, loosely dashed, very thick, red]
	table[x index=12, y index=13, 
	select coords between index={84}{89}, col sep = tab]
	{final_data.csv};
	\addplot+[mark=none, dotted, very thick, orange]
	table[x index=3, y index=4, 
	select coords between index={84}{89}, col sep = tab]
	{final_data.csv};
		\addplot+[mark=none, dashdotted, very thick, purple]
		table[x index=0, y index=1, 
	select coords between index={84}{89}, col sep = tab]
	{final_data.csv};
		\addplot+[mark=none, solid, very thick, black]
		table[x index=9, y index=10, 
	select coords between index={84}{89}, col sep = tab]
	{final_data.csv};
	\end{axis}
	\end{tikzpicture}
	\\
	
\emph{kodim13} & \emph{kodim14} & \emph{kodim15} \\[0.4cm]
	
		\begin{tikzpicture}
	\begin{axis}[height=0.3\textwidth, width=0.32\textwidth,
	xlabel={ {compression ratio}},
	ylabel={ {MSE}},
	grid = major
]
	\addplot+[mark=none, dashed, very thick, blue] 
	table[x index=6, y index=7, 
	select coords between index={90}{95}, col sep = tab] 
	{final_data.csv};
	\addplot+[mark=none, loosely dashed, very thick, red]
	table[x index=12, y index=13, 
	select coords between index={90}{95}, col sep = tab]
	{final_data.csv};
	\addplot+[mark=none, dotted, very thick, orange] 
	table[x index=3, y index=4, 
	select coords between index={90}{95}, col sep = tab]
	{final_data.csv};
		\addplot+[mark=none, dashdotted, very thick, purple]
		table[x index=0, y index=1, 
	select coords between index={90}{95}, col sep = tab]
	{final_data.csv};
		\addplot+[mark=none, solid, very thick, black]
		table[x index=9, y index=10, 
	select coords between index={90}{95}, col sep = tab]
	{final_data.csv};
	\end{axis}
	\end{tikzpicture} &
	
	\begin{tikzpicture}
	\begin{axis}[height=0.3\textwidth, width=0.32\textwidth,
	xlabel={ {compression ratio}},
	ylabel={ {MSE}},
	grid = major
]
	\addplot+[mark=none, dashed, very thick, blue] 
	table[x index=6, y index=7, 
	select coords between index={96}{101}, col sep = tab]
	{final_data.csv};
	\addplot+[mark=none, loosely dashed, very thick, red]
	table[x index=12, y index=13, 
	select coords between index={96}{101}, col sep = tab] 
	{final_data.csv};
	\addplot+[mark=none, dotted, very thick, orange] 
	table[x index=3, y index=4, 
	select coords between index={96}{101}, col sep = tab]
	{final_data.csv};
		\addplot+[mark=none, dashdotted, very thick, purple]
		table[x index=0, y index=1, 
	select coords between index={96}{101}, col sep = tab]
	{final_data.csv};
		\addplot+[mark=none, solid, very thick, black]
		table[x index=9, y index=10, 
	select coords between index={96}{101}, col sep = tab]
	{final_data.csv};
	\end{axis}
	\end{tikzpicture} & 
	\begin{tikzpicture}
	\begin{axis}[height=0.3\textwidth, width=0.32\textwidth,
	xlabel={ {compression ratio}},
	ylabel={ {MSE}},
	grid = major
]
	\addplot+[mark=none, dashed, very thick, blue]
	table[x index=6, y index=7, 
	select coords between index={102}{107}, col sep = tab]
	{final_data.csv};
	\addplot+[mark=none, loosely dashed, very thick, red]
	table[x index=12, y index=13, 
	select coords between index={102}{107}, col sep = tab]
	{final_data.csv};
	\addplot+[mark=none, dotted, very thick, orange]
	table[x index=3, y index=4, 
	select coords between index={102}{107}, col sep = tab]
	{final_data.csv};
		\addplot+[mark=none, dashdotted, very thick, purple] 
		table[x index=0, y index=1, 
	select coords between index={102}{107}, col sep = tab] 
	{final_data.csv};
		\addplot+[mark=none, solid, very thick, black] 
		table[x index=9, y index=10, 
	select coords between index={102}{107}, col sep = tab] 
	{final_data.csv};
	\end{axis}
	\end{tikzpicture}\\
	
	\emph{kodim16} & \emph{kodim17} & \emph{kodim18} \\[0.4cm]
	
			\begin{tikzpicture}
	\begin{axis}[height=0.3\textwidth, width=0.32\textwidth,
	xlabel={ {compression ratio}},
	ylabel={ {MSE}},
	grid = major
]
	\addplot+[mark=none, dashed, very thick, blue]
	table[x index=6, y index=7, 
	select coords between index={108}{113}, col sep = tab]
	{final_data.csv};
	\addplot+[mark=none, loosely dashed, very thick, red] 
	table[x index=12, y index=13, 
	select coords between index={108}{113}, col sep = tab]
	{final_data.csv};
	\addplot+[mark=none, dotted, very thick, orange]
	table[x index=3, y index=4, 
	select coords between index={108}{113}, col sep = tab]
	{final_data.csv};
		\addplot+[mark=none, dashdotted, very thick, purple] 
		table[x index=0, y index=1, 
	select coords between index={108}{113}, col sep = tab] 
	{final_data.csv};
		\addplot+[mark=none, solid, very thick, black]
		table[x index=9, y index=10, 
	select coords between index={108}{113}, col sep = tab]
	{final_data.csv};
	\end{axis}
	\end{tikzpicture} &
	
	\begin{tikzpicture}
	\begin{axis}[height=0.3\textwidth, width=0.32\textwidth,
	xlabel={ {compression ratio}},
	ylabel={ {MSE}},
	grid = major
]
	\addplot+[mark=none, dashed, very thick, blue] 
	table[x index=6, y index=7, 
	select coords between index={114}{119}, col sep = tab]
	{final_data.csv};
	\addplot+[mark=none, loosely dashed, very thick, red]
	table[x index=12, y index=13, 
	select coords between index={114}{119}, col sep = tab] 
	{final_data.csv};
	\addplot+[mark=none, dotted, very thick, orange]
	table[x index=3, y index=4, 
	select coords between index={114}{119}, col sep = tab] 
	{final_data.csv};
		\addplot+[mark=none, dashdotted, very thick, purple]
		table[x index=0, y index=1, 
	select coords between index={114}{119}, col sep = tab]
	{final_data.csv};
		\addplot+[mark=none, solid, very thick, black]
		table[x index=9, y index=10, 
	select coords between index={114}{119}, col sep = tab]
	{final_data.csv};
	\end{axis}
	\end{tikzpicture} & 
	\begin{tikzpicture}
	\begin{axis}[height=0.3\textwidth, width=0.32\textwidth,
	xlabel={ {compression ratio}},
	ylabel={ {MSE}},
	grid = major
]
	\addplot+[mark=none, dashed, very thick, blue]
	table[x index=6, y index=7, 
	select coords between index={120}{125}, col sep = tab]
	{final_data.csv};
	\addplot+[mark=none, loosely dashed, very thick, red]
	table[x index=12, y index=13, 
	select coords between index={120}{125}, col sep = tab]
	{final_data.csv};
	\addplot+[mark=none, dotted, very thick, orange] 
	table[x index=3, y index=4, 
	select coords between index={120}{125}, col sep = tab]
	{final_data.csv};
		\addplot+[mark=none, dashdotted, very thick, purple]
		table[x index=0, y index=1, 
	select coords between index={120}{125}, col sep = tab]
	{final_data.csv};
		\addplot+[mark=none, solid, very thick, black]
		table[x index=9, y index=10, 
	select coords between index={120}{125}, col sep = tab] 
	{final_data.csv};
	\end{axis}
	\end{tikzpicture}\\
	
\emph{kodim19} & \emph{kodim20} & \emph{kodim21} \\[0.4cm]
	
	\begin{tikzpicture}
	\begin{axis}[height=0.3\textwidth, width=0.32\textwidth,
	xlabel={ {compression ratio}},
	ylabel={ {MSE}},
	grid = major
]
	\addplot+[mark=none, dashed, very thick, blue]
	table[x index=6, y index=7, 
	select coords between index={126}{131}, col sep = tab]
	{final_data.csv};
	\addplot+[mark=none, loosely dashed, very thick, red]
	table[x index=12, y index=13, 
	select coords between index={126}{131}, col sep = tab] 
	{final_data.csv};
	\addplot+[mark=none, dotted, very thick, orange]
	table[x index=3, y index=4, 
	select coords between index={126}{131}, col sep = tab]
	{final_data.csv};
		\addplot+[mark=none, dashdotted, very thick, purple]
		table[x index=0, y index=1, 
	select coords between index={126}{131}, col sep = tab]
	{final_data.csv};
		\addplot+[mark=none, solid, very thick, black] 
		table[x index=9, y index=10, 
	select coords between index={126}{131}, col sep = tab]
	{final_data.csv};
	\end{axis}
	\end{tikzpicture} &
	
	\begin{tikzpicture}
	\begin{axis}[height=0.3\textwidth, width=0.32\textwidth,
	xlabel={ {compression ratio}},
	ylabel={ {MSE}},
	grid = major
]
	\addplot+[mark=none, dashed, very thick, blue]
	table[x index=6, y index=7, 
	select coords between index={132}{137}, col sep = tab] 
	{final_data.csv};
	\addplot+[mark=none, loosely dashed, very thick, red]
	table[x index=12, y index=13, 
	select coords between index={132}{137}, col sep = tab] 
	{final_data.csv};
	\addplot+[mark=none, dotted, very thick, orange]
	table[x index=3, y index=4, 
	select coords between index={132}{137}, col sep = tab]
	{final_data.csv};
		\addplot+[mark=none, dashdotted, very thick, purple] 
		table[x index=0, y index=1, 
	select coords between index={132}{137}, col sep = tab] 
	{final_data.csv};
		\addplot+[mark=none, solid, very thick, black]
		table[x index=9, y index=10, 
	select coords between index={132}{137}, col sep = tab] 
	{final_data.csv};
	\end{axis}
	\end{tikzpicture} & 
	\begin{tikzpicture}
	\begin{axis}[height=0.3\textwidth, width=0.32\textwidth,
	xlabel={ {compression ratio}},
	ylabel={ {MSE}},
	grid = major
]
	\addplot+[mark=none, dashed, very thick, blue] 
	table[x index=6, y index=7, 
	select coords between index={138}{143}, col sep = tab] 
	{final_data.csv};
	\addplot+[mark=none, loosely dashed, very thick, red]
	table[x index=12, y index=13, 
	select coords between index={138}{143}, col sep = tab]
	{final_data.csv};
	\addplot+[mark=none, dotted, very thick, orange] 
	table[x index=3, y index=4, 
	select coords between index={138}{143}, col sep = tab]
	{final_data.csv};
		\addplot+[mark=none, dashdotted, very thick, purple]
		table[x index=0, y index=1, 
	select coords between index={138}{143}, col sep = tab]
	{final_data.csv};
		\addplot+[mark=none, solid, very thick, black]
		table[x index=9, y index=10, 
	select coords between index={138}{143}, col sep = tab]
	{final_data.csv};
	\end{axis}
	\end{tikzpicture}\\
	
	\emph{kodim22} & \emph{kodim23} & \emph{kodim24} \\[0.4cm]

\end{tabular}
\caption{Rate-distortion comparisons on the Kodak database}
		
\end{figure*}

\end{document}